\documentclass[a4paper,11pt]{article}
\pdfoutput=1 

\usepackage{jcappub} 
\usepackage[T1]{fontenc} 

\usepackage{natbib}
\usepackage{cancel}
\usepackage{enumitem}    
\usepackage{amsthm}         
\usepackage{amsmath} 	   
\usepackage{amssymb}       
\usepackage{amsfonts}
\usepackage{graphicx} 
\usepackage{dblfloatfix}	    
\usepackage{verbatim}
\usepackage{epsfig,multicol,bbm}
\usepackage{color}
\usepackage{colortbl}
\usepackage{float}
\usepackage{multirow}
\usepackage{array}

\usepackage{subfig}

\allowdisplaybreaks

\title{Robustness of slow contraction to cosmic initial conditions}
\author[a,b,1]{Anna Ijjas,}
\note{Corresponding Author}
\author[c]{William G. Cook,}
\author[d]{Frans Pretorius,}
\author[d]{Paul J. Steinhardt,}
\author[d]{and Elliot Y. Davies}

\affiliation[a]{Max Planck Institute for Gravitational Physics (Albert Einstein Institute), D-30167 Hannover, Germany}
\affiliation[b]{Gottfried Wilhelm Leibniz Universit\"at, D-30167 Hannover, Germany}
\affiliation[c]{Theoretisch-Physikalisches Institut, Friedrich-Schiller-Universit\"at, D-07743 Jena, Germany}
\affiliation[d]{Department of Physics, Princeton University, Princeton, NJ 08544, USA}

\emailAdd{anna.ijjas@aei.mpg.de}

\abstract{
We present numerical relativity simulations of cosmological scenarios in which the universe is smoothed and flattened by undergoing a phase of slow contraction and test their sensitivity to a wide range of initial conditions. Our numerical scheme enables the variation of all freely specifiable physical quantities that characterize the initial spatial hypersurface, such as the initial shear and spatial curvature contributions as well as the initial field and velocity distributions of the scalar that drives the cosmological evolution.
In particular, we  include initial conditions that are far outside the perturbative regime of the well-known attractor scaling solution.
We complement our numerical results by analytically performing a complete dynamical systems analysis and show that the two approaches yield consistent results.
}


\begin{document}
\maketitle 
\raggedbottom

\section{Introduction}
\label{sec_intro}
The cosmic initial conditions problem presents an unusual theoretical puzzle. Typically, in understanding a dynamical system, the challenge lies in finding the right (differential) equations that describe how the system evolves with time. The specific point at which one starts solving the evolution equations is of no particular importance, especially because the intent is to apply the same equations  to understanding the behavior over a wide range of systems and initial conditions. 
In cosmology, on the other hand, the relevant dynamical equations are well-known.
Measurements of the cosmic microwave background and other astrophysical experiments confirm that Einstein's classical theory of general relativity describes the time evolution of our large-scale universe since primordial nucleosynthesis to an astonishing accuracy, {\it if} we specify the initial geometry and radiation-matter content. The very same observations, though, also teach us that the cosmic initial conditions cannot be explained by the same physics that underlies the evolution equations  because, in that context, they are exponentially rare or finely tuned. Some mechanism operating before primordial nucleosynthesis is needed to explain how these initial conditions arise.

The goal of this paper is to show that slow contraction is a `robust' smoothing mechanism that can naturally generate the cosmic initial conditions.  The slow contraction phase is induced by a canonical scalar field evolving down a steep negative potential coupled to Einstein general relativity, as postulated in many bouncing and cyclic cosmological models \cite{Khoury:2001wf,Steinhardt:2001st,Ijjas:2018qbo,Ijjas:2019pyf}.  We adapt the tools of numerical general relativity to solve the full set of coupled Einstein-scalar field equations beginning from a wide range of highly inhomogeneous and anisotropic states that would not be compatible with observations of the cosmic microwave background and large-scale structure if there were not a robust smoothing and flattening mechanism.  The measure of `robustness' is how sensitive the outcome is to   the initial state and whether the outcome converges closely and rapidly to the initial conditions needed to explain cosmological observations.  

More precisely, the large-scale universe evolves as predicted by the Einstein equations {\it if } the cosmic initial conditions at primordial nucleosynthesis are very precisely set as follows:
\begin{enumerate}
\item[i.] the background spacetime geometry is smooth and spatially flat described by a single dynamical quantity, the Friedmann-Robertson-Walker (FRW)  scale factor $a(\tau)$, or equivalently, the associated Hubble radius $\Theta=|H^{-1}|$.   (Here and throughout, the Hubble parameter $H$ is defined as the logarithmic time derivative of the scale factor $H = a^{\prime}/a$ and prime denotes differentiation with respect to the proper FRW time coordinate $\tau$.);
\item[ii.] to first sub-leading order, there is only a single kind of deviation from this geometry, namely a nearly scale-invariant and gaussian spectrum of adiabatic curvature fluctuations;
\item[iii.] there are no measurable first-order tensor, vector, or isocurvature fluctuations in the initial space-time geometry;  
\item[iv.] the initial background geometry as well as the initial spectrum of curvature fluctuations are correlated over $10^{29}$ (or, equivalently, $e^{60}$) Hubble-sized patches.
\end{enumerate} 
Notably, under these initial conditions, the Einstein equations dramatically simplify and the subsequent large-scale evolution of 14 billion years  is given by the Friedmann equation,
\begin{equation}
\label{frw1}
\frac{{\rm d} H^{-1}}{{\rm d}\tau}=\varepsilon_{\rm eff},
\end{equation}
where $\varepsilon_{\rm eff}\equiv (3/2)(1+p/\varrho)$ is the equation of state of the dominant stress-energy component, relating its pressure $p$ to its energy density $\varrho$. (Throughout, we work in reduced Planck units.)
Furthermore, as the background slowly expands, the initial curvature fluctuations in the space-time geometry grow and become the seeds from which galaxies, stars and planets form. 

However, in a classical general relativistic space-time, this combination of initial conditions is highly non-generic if the stress-energy is sourced by ordinary (baryonic or dark) matter and radiation. Accordingly, the cosmic initial conditions problem consists in identifying a mechanism that {\it generically} (i.e., for most initial data) leads to the conditions as given through (i.)-(iv.).

The traditional approach to resolving the cosmic initial conditions problem has been to propose a `{\it classical}' smoothing mechanism \cite{Cook:2020oaj}. For this purpose, it is necessary that the mechanism possesses a dynamical attractor solution in which the relative contribution of {\it small} inhomogeneities and anisotropies to the total energy density shrinks with time. 
In all known examples, classical smoothing is achieved by introducing a novel stress-energy component, typically sourced by scalar field $\phi$,  that evolves to dominate all other contributions to the generalized Friedmann constraint,
\begin{equation}
\label{frw-const}
H^2 = \frac13\left( \frac{\varrho_m}{a^3} + \frac{\varrho_r}{a^4} + \frac{\varrho_{\phi}}{a^{2\varepsilon}} \right) -\frac{k}{a^2} +\frac{\sigma^2}{a^6}.
\end{equation}
Here, $\varrho_m, \varrho_r, \varrho_{\phi}$ represent the energy density of matter, radiation and the scalar field $\phi$, respectively, at some initial time $t_0$,  and the scale factor is normalized such that $a(t_0) = 1$. The last two terms correspond to spatial curvature and anisotropy.

From Eq.~\eqref{frw-const}, it is immediately clear that there can only be two kinds of classical smoothing mechanisms: 
\begin{itemize}
\item in an expanding universe ($H>0$): $0\leq\varepsilon<1$ ({\it inflation}); 
\item in a contracting universe ($H<0$): $\varepsilon>3$ ({\it slow contraction}). 
\end{itemize}
The respective equation of state $\varepsilon$ is achieved by choosing a suitable potential energy density $V(\phi)$. For example, assuming a scalar field with canonical kinetic energy density and a negative exponential potential 
\begin{equation}
\label{potential}
V(\phi) = V_0 \exp(-\phi/M),
\end{equation}
 where $V_0<0$, the contracting FRW space-time is stable to small perturbations and hence a dynamical attractor solution of the Einstein-scalar field equations, if the characteristic energy scale $M<\sqrt{6}$. In this case, $M^{-1}=\sqrt{2\varepsilon}$. 

In addition,  both inflationary expansion and slow contraction transform quantum fluctuations generated on scales smaller than the Hubble radius into  `squeezed modes' on scales larger than the Hubble radius, as required by condition (ii). During slow contraction, for example, the characteristic wavelength of a fluctuation, $\lambda$, decreases in proportion to the scale factor, $\lambda \rightarrow \lambda \times a(t)$; but the Hubble radius decreases more rapidly, as  $\Theta \sim a^{\varepsilon}$ where $\varepsilon >3$. Consequently, a mode that originates on scales much smaller than a Hubble patch evolves to a wavelength extending over scales exponentially larger than the Hubble radius.

Yet, classical smoothing is necessary but not sufficient by far to explain our large-scale universe. Satisfying this criterion alone would mean that only an infinitesimal set of initial conditions, namely small classical perturbations around FRW, lead to a universe as we observe it.
Another, obvious requirement is that the classical evolution remains stable to quantum fluctuations. Notably, inflation is a classical but famously {\it not} a quantum smoother \cite{Steinhardt:1982kg,Vilenkin:1983xq,Guth:2007ng};  but slow contraction is {\it both} a classical {\it and} a quantum smoother \cite{Cook:2020oaj}. 

More importantly, a classical smoothing phase, even if it is stable to quantum fluctuations,  does not solve the cosmic initial conditions problem if it is only stable under small perturbations around a smooth and flat FRW background. After all, the set of space-time geometries that represent small deviations from a FRW space-time only represent a measure-zero set of all permitted and (physically) plausible initial conditions. 
To {\it generically} drive the universe towards a flat, homogeneous and anisotropic space-time, a classical smoothing mechanism must be a `{\it robust}' smoother, {\it i.e.}, insensitive to a wide range of arbitrary initial conditions including those well outside the perturbative regime of the attractor FRW solution.

In this paper, we examine quantitatively the robustness of slow contraction using a numerical scheme that enables the variation of all freely specifiable physical quantities that characterize the initial spatial hypersurface, such as the initial shear and spatial curvature contributions as well as the initial field and velocity distributions of the scalar that drives the cosmological evolution. In fact, the only restrictions we have on the initial data result from imposing periodic spatial boundary conditions and choosing an initial three-metric that is conformally flat.  

In particular, we `empirically' confirm the well-known ultra-locality conjecture \cite{Belinsky:1970ew} and demonstrate that, generically, all gradients rapidly become negligible as the evolution proceeds.
Finally, we show that the homogenous end states we identified numerically are the only stable attractor solutions of the underlying relativistic system of evolution and constraint equations.

\section{Evolution and constraint equations in orthonormal tetrad form}
\label{sec_methods}

To study the robustness of slow contraction in smoothing and flattening the universe, we solve the full Einstein-scalar field equations for a wide range of highly non-perturbative initial conditions using the techniques of numerical general relativity.
The study entails numerically evolving a system of coupled non-linear, second-order partial differential equations.
Performing such a computation necessitates finding a `good' formulation of the field equations satisfying two criteria:
\begin{itemize}
\item the formulation is `well-suited' to the physical situation; and
\item the formulation is well-posed.
\end{itemize}
The term `formulation' is given multiple definitions in the literature. Here, we follow the convention established in mathematical relativity ({\it e.g.}, see \cite{Geroch-IVP}): A `formulation' (sometimes called an initial value formulation) of a given theory is the representation of the underlying system of differential equations obtained by choosing a particular coordinate system.  
Furthermore, we define a formulation to be `well-posed'  if  the underlying system of differential equations can be put into a form such that, for given initial conditions, there exists a unique solution that depends continuously on the initial conditions.  Note that a formulation is not equivalent to a gauge choice. Different gauge conditions can be implemented in the same formulation, but not all gauge choices lead to a well-posed formulation.\footnote{We note that some fields use the phrase `well-posed problem'
as we use `well-posed formulation' here. The latter is a common usage in the modern relativity literature and implicitly includes that appropriate boundary conditions and initial data are specified.}

Typical cosmological studies in the literature are {\it perturbative} and only consider the second criterion.
The diffeomorphism invariance of the field equations is exploited to simplify the theoretical analysis and straightforwardly relate the predictions of different cosmological models to observations. 

For example, in studying the evolution of perturbations away from FRW in the very early universe, 
a coordinate basis using the well-known $(3+1)$ or Arnowitt-Deser-Misner (ADM) decomposition \cite{Arnowitt:1959ah} proves to be well-suited. The ADM formulation naturally rests on the homogeneity and isotropy of the FRW background solution, enabling the separation of linear perturbations around this background into decoupled scalar, vector, and tensor degrees of freedom that each evolve independently mode by mode \cite{Lifshitz:1945du,Gerlach:1979rw,Bardeen:1980kt}. 
Finally, a suitable gauge choice is made for the physical situation at hand. 
In a gauge such as unitary,  the scalar part of the linearized spatial metric component can be identified with the co-moving curvature perturbation, an invariant of the linear theory that  directly determines the temperature anisotropies of the cosmic microwave background (CMB) \cite{Bardeen:1983qw,Mukhanov:1988jd}.

However, for the {\it non-perturbative} numerical analyses presented in this paper, the common procedure adapted in the cosmology literature is insufficient.  Without  a well-posed formulation of the full (non-perturbative) Einstein-scalar field equations, the existence of a unique solution is not guaranteed.  For a given set of initial conditions, the system of equations might admit no solution at all or it might admit multiple solutions.  That is, in an ill-posed formulation, no predictions can be derived.  A common manifestation is the `blow up' of the numerical code within finite time even if there is no fundamental instability in the underlying theory.

Notably, many formulations of the Einstein-scalar field equations, including the ADM decomposition popular with cosmologists, are {\it ill-posed}.  In particular, for the case where the lapse and shift are given by algebraic relations, it is straightforward to show that the resulting partial differential equations for the evolution subsystem of the Einstein-scalar field system is only weakly hyperbolic rather than well-posed (strongly hyperbolic).
Strictly speaking, cosmologists focusing on perturbative cosmological analyses are only able to get by with an ADM decomposition  because there exist other formulations of Einstein scalar-field equations that are well-posed and that have been shown to uniquely admit the FRW solution assumed in ADM.

In general, we must  not only choose a well-posed formulation of the field equations but, in some cases, one that also admits an effective constraint damping scheme.  Constraint damping is a common tool used in numerical relativity. In principle, by choosing initial conditions that satisfy the Hamiltonian and momentum constraints, {\it i.e.}, the Einstein equations projected orthogonally onto a spacelike hypersurface at some initial time $t_0$, the field equations of general relativity propagate and preserve the constraints going forward in time. In practice, though, the constraints can only be satisfied initially up to some numerical error, and, in some cases, the error  grows exponentially.  Then, even a well-posed formulation can result in a numerical blow-up if these constraint violations are not addressed. For this reason, in some numerical setups it is often prudent
to add terms to the evolution equations that damp the growth of constraint violating numerical errors, while leaving the underlying solution unaffected in the continuum limit \cite{Pretorius:2004jg,Pretorius:2006tp}. As it turns out, in the numerical scheme described in this paper, constraint damping is not required for stability of the numerical evolution.

In the cases considered in this paper, we need a `suitable' formulation of the equations that address  issues specifically related to slowly contracting spacetimes: a stiffness problem and accurately tracking contraction over many e-folds before reaching a big crunch. 
The stiffness problem arises because the equation of state parameter $\varepsilon$ is greater than three (with typical models having  $ \varepsilon \gg 3$) which means that  $\Theta \propto a^{\varepsilon}$  is {\it very rapidly} decreasing
 compared to $a(t)$.
For example, in models discussed in the literature \cite{Ijjas:2019pyf}, during a period when the averaged scale factor decreases a factor of two or three,  the Hubble radius  shrinks by a factor of $e^{120}$ or more.  
To handle this stiffness problem, we choose a {\it Hubble-normalized} formulation in which the Hubble radius $\Theta$ does not enter explicitly.   

To handle the crunch problem, 
we  choose a time coordinate such that it follows the mean curvature growth, {\it i.e.}  a time slicing where $ e^{t} \equiv 3 \Theta $.    This way, reaching the curvature singularity (or vanishing of the Hubble radius $\Theta$)  takes an infinite (coordinate) time.  We find that our Hubble-normalized formulation using this time slicing  is sufficiently suitable (our third criterion above) for analyzing slow-contraction.

We find that an effective way of incorporating these methods of handling the stiffness and crunch problems within a well-posed initial value formulation is to adapt the orthonormal tetrad formulation of the Einstein-scalar field equations in our numerical scheme. Originally, the formulation was used by Sch\"ucking to find all exact vacuum solutions describing spatially homogeneous spacetimes \cite{2003GReGr..35..491K,Estabrook:1964zk}. Later, it was successfully implemented to numerically studying contracting vacuum spacetimes \cite{Estabrook:1996wa,Buchman:2003sq,Bardeen:2011ip} as well as spacetimes with a canonical scalar field \cite{Garfinkle:2008ei}.  In Sec.~\ref{subsec_var}, we first introduce the basic tetrad variables. Then, we derive the Einstein scalar field equations in tetrad form in Sec.~\ref{subsec_tetrad-eqs}. Finally, in Sec.~\ref{subsec_num-eqs}, we convert the tetrad equations to partial differential equations using local coordinates, making the system readily usable as a numerical evolution scheme.

\subsection{Variables}
\label{subsec_var}

Tetrad formulations of the Einstein-scalar field equations assign each spacetime point a family of unit basis 4-vectors (or {\it vierbeins}) $\{e_0, e_1, e_2, e_3\}$ (as opposed to coordinates $\{x_0, x_1, x_2, x_3\}$) that describe local Lorentz frames with the spacetime metric being given by the dot product of the basis vectors. The starting point is a timelike vector field $e_0$ that defines a future-directed  timelike reference congruence, to which it is tangent. It is supplemented by a triad of spacelike unit 4-vectors $\{e_1, e_2, e_3\}$ that lie in the rest 3-spaces of $e_0$. 

In an {\it orthonormal} tetrad formulation that we shall employ, the spacetime metric is everywhere given by
\begin{equation}
 e_{\alpha} \cdot e_{\beta} = \eta_{\alpha\beta},
\end{equation}
where $\eta_{\alpha\beta}={\rm diag}(-1,1,1,1)$ the Minkowski metric and "$\cdot$" is the spacetime inner product. 
Throughout, spacetime indices $(0-3)$ are Greek and spatial indices (1-3) are Latin. The beginning of the alphabet ($\alpha, \beta, \gamma$ or $a,b,c$) is used for tetrad indices and the middle of the alphabet  ($\mu, \nu, \rho$ or $i,j,k$) is used for coordinate indices. Tetrad frame indices are raised and lowered with $\eta_{\alpha \beta}$.

The geometric variables of the formulation are the sixteen tetrad vector components and the twenty-four Ricci rotation coefficients
\begin{equation}
\label{RicRot}
\gamma_{\alpha\beta\gamma} \equiv e_{\alpha} \cdot \nabla_{\gamma} e_{\beta} = - \gamma_{\beta\alpha \gamma},
\end{equation}
which  define the deformation of the tetrad when moving from point to point.  Here, $\nabla_{\gamma}$ is the spacetime covariant derivative projected onto a tetrad $e_{\gamma}{}^{\lambda}\,\nabla_{\lambda}$. Note that the $\gamma_{\alpha\beta\gamma}$ are the `tetrad components' of the Christoffel symbols $\Gamma_{\mu\nu\rho}$.

Similar to coordinate-based formulations of the field equations, the tetrad formulation greatly simplifies when making a space-time split. Unlike in the 3+1 (coordinate based) ADM formulation, where the split is defined by the constant-time spacelike hypersurface, here the split is relative to the timelike congruence defined by $e_0$. Note, though, that the timelike congruence does {\it not} uniquely define the auxiliary spatial congruence. Rather, the spatial triad vectors are fixed by imposing gauge conditions.

To perform the spacetime split, we first write the fifteen connection coefficients that have at least one timelike index in terms of 3-dimensional quantities $b_a, \Omega_a$, and $K_{ab}$, reflecting the antisymmetry of $\gamma_{\alpha\beta\gamma}$ in its first two indices:
\begin{eqnarray}
\gamma_{a00} &=& - \gamma_{0a0} =  b_a
,\\
\gamma_{ab0} &=& - \gamma_{ba0} = \epsilon_{abc}\Omega^c
,\\
 \gamma_{0ab} &=&  - \gamma_{a0b} = -K_{ba} 
;
\end{eqnarray}
where $\epsilon_{abc}$ is the Levi-Civita symbol. 
As shown by Ehlers et al. \cite{1961MAWMN..11..792E,2009GReGr..41.2191J}, the fifteen quantities describe kinematic fields associated with the timelike congruence tangent to $e_0$: the 3-vector $b_a$ is the local proper acceleration; the 3-vector $ \Omega_a$ is the local angular velocity of the space-like triads $\{e_1, e_2, e_3\}$ relative to Fermi-propagated axes; and $K_{ba}$ is the local rate-of-strain (or shear) tensor. (A spatial triad is `Fermi-propagated' if $ \Omega_a\equiv0$, {\it i.e.}, it is a local, inertially non-rotating reference frame.)
 
Similarly, we express the remaining nine purely spatial connection coefficients $\gamma_{abc}$ that describe the induced curvature of the auxiliary 3-congruence using a 3-tensor,
\begin{equation}
N_{ab} = \textstyle{\frac12}\epsilon_{b}{}^{cd} \gamma_{cda}.
\end{equation}
The spatial connection coefficients $N_{ab}$ and the components of the shear tensor $K_{ab}$ are the eighteen dynamical variables. The acceleration and angular velocity vectors, $b_a, \Omega_a$, are the six (tetrad) gauge source functions.

For completeness, we note that some authors use as basic variables the commutation (or structure) coefficients ${\cal C}_{\alpha\beta\gamma}$ rather than the Ricci rotation coefficients $\gamma_{\alpha\beta\gamma}$ \cite{Ellis:1966ta,vanElst:1996dr,vanElst:2001xm}. 
Here, the ${\cal C}_{\alpha\beta\gamma}$ are given through
\begin{equation}
\label{comm}
[e_{\alpha}, e_{\beta}] \equiv  \nabla_{\alpha} e_{\beta} - \nabla_{\beta} e_{\alpha} \equiv {\cal C}_{\alpha\beta\gamma} e^{\gamma}.
\end{equation}
It is straightforward to relate the two conventions: With the definitions in Eqs.~\eqref{RicRot} and \eqref{comm},
\begin{equation}
\gamma_{\alpha\beta\gamma} = \frac12\Big({\cal C}_{\gamma\beta \alpha} + {\cal C}_{\alpha\gamma\beta } -{\cal C}_{\beta\alpha\gamma}\Big), 
\end{equation} 
or, alternatively, 
\begin{equation}
{\cal C}_{\alpha\beta\gamma} = \gamma_{\gamma\beta \alpha} - \gamma_{\gamma\alpha\beta}, 
\end{equation} 
{\it i.e.}, expressed in terms of the twenty-four 3D quantities, the full set of commutation coefficients takes the following form,
\begin{eqnarray}
\label{comm-1}
{\cal C}_{a00} &=& - {\cal C}_{0a0} = b _a
,\\
{\cal C}_{ab0} &=& - {\cal C}_{ba0} = 2 \epsilon_{abc}\omega^c
,\\
 {\cal C}_{0ab} &=& -{\cal C}_{a0b} =  -K_{(ab)} + \epsilon_{abc}\left( \omega^c -\Omega^c \right)
,\\
\label{comm-4}
{\cal C}_{abc} &=& - {\cal C}_{bac} = \epsilon_{cb}{}^d N_{ad} 
.
\end{eqnarray}
Here, the 3-vector $\omega_a$ is the antisymmetric part of $K_{ab}$,
\begin{equation}
\omega_a \equiv {\textstyle \frac12} \epsilon_{a}{}^{bc}K_{bc};
\end{equation}
it measures the vorticity (or twist) vector of the $e_0$-congruence. 

Finally, the geometric variables have to be supplemented by the dynamical quantities that describe the matter source. In our case, this is the canonical scalar field $\phi$ which is specified in the Einstein-scalar field equations through its potential energy density $V(\phi)$.

\subsection{Tetrad equations}
\label{subsec_tetrad-eqs}

Next we present the tetrad evolution and constraint equations for the (eighteen) dynamical variables $K_{ab}, N_{ab}$. The remaining (six) gauge variables are fixed by our tetrad frame gauge choice.

A natural gauge choice is a frame with 
\begin{enumerate}
\item[i.]Fermi-propagated axes ($\Omega_a\equiv0$); and 
\item[ii.] hypersurface orthogonal timelike congruence ($\omega_a\equiv0$ or, equivalently, $K_{ab}\equiv K_{(ab)}$). 
\end{enumerate}
Here and throughout, parentheses denote symmetrization, {\it i.e.}, $K_{(ab)}\equiv {\textstyle \frac12} (K_{ab}+K_{ba})$.
In this (frame) gauge, the time-like {\it vierbein} $e_0$ is the future-directed unit normal to the spacelike hypersurfaces $\Sigma_t$ of constant time, and the spatial tetrad vectors are tangent to $\{\Sigma_t\}$. Furthermore, $K_{ab}$ is the extrinsic curvature of $\Sigma_t$ and the $N_{ab}$ are the nine (intrinsic) spatial curvature variables. All Ricci rotation coefficients, $K_{ab}, N_{ab}$ act as scalars  on $\Sigma_t$.

Employing these gauge conditions, the tetrad evolution and constraint equations take the following form:
\begin{eqnarray}
\label{eq-K-ab}
D_0 K_{ab} &=& \epsilon_{a}{}^{cd}D_c N_{db} + D_a b_b+ b_a b_b - \epsilon_{b}{}^{cd}N_{ac}b_d + N_c{}^cN_{ab}  -K_{a}{}^{c}K_{cb} - N_{ca}N^c{}_{b} \\
&+& {\textstyle \frac12} \epsilon_{a}{}^{df}\epsilon_{b}{}^{ce}\left( K_{dc}K_{fe} - N_{dc}N_{fe} \right)
+ s_{ab} -  {\textstyle \frac12}\delta_{ab}\big(\varrho + 3p\big)
,\nonumber\\
\label{eq-N-ab}
D_0 N_{ab} &=& - \epsilon_{a}{}^{cd}D_c K_{db} +  \epsilon_{b}{}^{cd}K_{ac}b_d - N_c{}^c K_{ab} 
+ 2 N_{c[a}K_{b]}{}^c + \epsilon_{a}{}^{df}\epsilon_{b}{}^{ce}N_{dc}K_{fe}\\
&-& \epsilon_{ab}{}^{c}\,j_c,
\nonumber\\
\label{H-const}
2D_b A^b&=&N_{ab}N^{ab} + {\textstyle \frac12}\left( K_{ab}K^{ab} - N_{ab}N^{ab} - (K_a{}^a)^2 - (N_a{}^a)^2\right) + \varrho
,\\
\label{m-const}
D_b K_a{}^b &-& D_a K_c{}^c = \epsilon_{a}{}^{bc}K_b{}^d N_{dc} + 2 K_a{}^c A_c - j_a
,\\
\label{N-ab-const}
D_b N_{a}{}^b &-& D_a N_c{}^c = - \epsilon_{a}{}^{bc}N_b{}^d N_{dc} 
,
\end{eqnarray}
where 
\begin{equation}
A_b \equiv {\textstyle \frac12} \epsilon_{b}{}^{cd}N_{cd}.
\end{equation}
is the antisymmetric part of $N_{ab}$; $D_0$ is the Lie derivative along the timelike {\it vierbein} $e_0$; and $D_a$ denotes the directional derivative along the spatial {\it vierbein} $e_a$. 
The matter variables associated with the stress-energy $T_{\alpha\beta}$, such as the energy density $\varrho$, pressure $p$, 3-momentum flux $j_a$, and (spatial) stress tensor $s_{ab}$, are defined as follows:
\begin{eqnarray} 
\varrho &\equiv& e_0{}^{\alpha}e_0{}^{\beta}T_{\alpha\beta}
,\\ 
j_a &\equiv& - e_0{}^{\alpha}e_a{}^{\beta}T_{\alpha\beta},\\
s_{ab} &\equiv& e_a{}^{\alpha}e_b{}^{\beta}T_{\alpha\beta},\\ 
p &\equiv& {\textstyle \frac13} s_c{}^c.
\end{eqnarray}
Notably, there is no evolution equation for the acceleration 3-vector $b_a$, which reflects the fact that it is a (frame) gauge source function.  

The tetrad evolution and constraint equations (\ref{eq-K-ab}-\ref{N-ab-const}) were first obtained in Ref.~\cite{Buchman:2003sq}. Their derivation is straightforward when using the tetrad form of the Riemann tensor,
\begin{equation}
\label{riemann-def}
R_{\alpha\beta\gamma\delta} = D_{\gamma}\gamma_{\alpha\beta\delta} - D_{\delta}\gamma_{\alpha\beta\gamma} + \gamma_{\alpha\epsilon\gamma}\gamma^{\epsilon}{}_{\beta\delta}
- \gamma_{\alpha\epsilon\delta}\gamma^{\epsilon}{}_{\beta\gamma}
+ \gamma_{\alpha\beta\epsilon}\left( \gamma^{\epsilon}{}_{\gamma\delta}- \gamma^{\epsilon}{}_{\delta\gamma} \right)
.
\end{equation}
More exactly, substituting Eq.~\eqref{riemann-def} into the (trace-reversed) Einstein-scalar field equations,
\begin{equation}
R_{\alpha\beta} = T_{\alpha\beta} - {\textstyle \frac12} \eta_{\alpha\beta} \eta^{\gamma\delta}T_{\gamma\delta},
\end{equation}
where $R_{\alpha\beta}  \equiv R^{\gamma}{}_{\alpha\gamma\beta}$ is the Ricci tensor,
yields the evolution equation~\eqref{eq-K-ab} for $K_{ab}$ as well as the Hamiltonian and momentum constraints, Eqs.~\eqref{H-const} and \eqref{m-const}, respectively.
The evolution and constraint equations for the spatial curvature variables $N_{ab}$, Eqs.~(\ref{eq-N-ab}, \ref{N-ab-const}), follow from the Riemann identities,
\begin{align}
&R_{\alpha\beta\gamma\delta} = R_{\gamma\delta\alpha\beta}
,\\
&R_{\alpha\beta\gamma\delta} + R_{\alpha\delta\beta\gamma} + R_{\alpha\gamma\delta\beta} =0.
\end{align}

For a single scalar field with canonical kinetic energy density and (non-zero) potential $V(\phi)$,
the hydrodynamical (macroscopic) matter variables $\varrho, p, j_a$ and $s_{ab}$ are given by
\begin{eqnarray}
\varrho &=& {\textstyle \frac12}D_0\phi D_0\phi + {\textstyle \frac12}D^a\phi D_a\phi + V(\phi),\\
s_{ab} &=& D_a\phi D_b\phi + \left( {\textstyle \frac12}D_0\phi D_0\phi - D_c\phi D^c\phi-V(\phi)\right)\delta_{ab}
,\\
p &=&  {\textstyle \frac12} D_0\phi D_0\phi - {\textstyle \frac16}D^a\phi D_a\phi -  V(\phi)
,\\
j_a &=& - D_0\phi D_a\phi 
.
\end{eqnarray}
Note that, in general, $j_a$  is non-zero and $s_{ab}$ is non-diagonal, which reflects the fact that choosing a hypersurface-orthogonal tetrad frame gauge generally does {\it not} coincide with the co-moving frame of the (scalar field) matter source. In fact, a hypersurface-orthogonal frame-gauge is co-moving only in the homogeneous (ultra-local) limit.

The system~(\ref{eq-K-ab}-\ref{N-ab-const}) is completed by adding the scalar-field evolution and constraint equations:
\begin{eqnarray}
\label{eq-phi}
D_0 \phi &=& W
,\\
\label{const-phi}
D_a \phi &=& S_a
,\\
\label{eq-W}
D_0 W &=& -\delta^{ab}K_{ab}W + D_a S^a + \left(b_a - 2A_a \right) S^a- V,_{\phi}
,\\
\label{eq-S-a}
D_0 S_a &=& D_a W + b_a W - K_{(ab)}S^b
.
\end{eqnarray}
Here, Eqs.~\eqref{eq-phi} and \eqref{const-phi} are the defining relations for the auxiliary variables $W$ and $S_a$, which denote the velocity and gradient of the scalar field $\phi$, respectively. Eq.~\eqref{eq-W} is obtained by expanding the Laplacian of the Klein-Gordon equation ($\Box\phi=V,_{\phi}$) using the Ricci rotation coefficients; and Eq.~\eqref{eq-S-a} is obtained by evaluating the commutation relation $[e_0, e_a]\phi = -{\cal C}_{0a0}W + {\cal C}_{0ab}S^b $ using Eqs.~(\ref{comm-1}-\ref{comm-4}).

%
%

\subsection{Numerical evolution scheme}
\label{subsec_num-eqs}

In order to evolve the tetrad equations~(\ref{eq-K-ab}-\ref{N-ab-const}, \ref{eq-phi}-\ref{eq-S-a}) numerically, we must write them as a system of partial differential equations. That means, we must give a representation of the tetrad vector components $\{e_{\alpha}\}$ using a particular set of local coordinates $\{x^{\mu}\}$ and then convert the directional derivatives $D_{\alpha}$ in the tetrad equations to partial derivatives along these coordinates.  

To this end, we introduce the transformation matrix $\{\lambda_{\alpha}^{\mu}\}$ between coordinate and tetrad basis vectors defined thru
\begin{equation}
e_{\alpha} = \lambda_{\alpha}^{\mu} e_{\mu}
.
\end{equation}
The coordinate metric components are then given by 
\begin{equation}
\label{g-mu-nu-rel}
g^{\mu\nu} = \eta^{\alpha\beta} \lambda_{\alpha}^{\mu}\lambda_{\beta}^{\nu};
\end{equation}
and
directional derivatives along tetrads can now be written as partial derivatives along coordinate directions,
 \begin{equation}
D_0 = N^{-1}\left( \partial_t - N^i\partial_i \right),\quad D_a = E_a{}^i \partial_i,
\end{equation}
where $N$ is the tetrad lapse function and the $N^i$ are the three coordinate components of the tetrad shift vector. Both the tetrad lapse function and the tetrad shift vector describe the evolution of the coordinates relative to the tetrad congruence (as opposed to the ADM lapse and shift that describe the evolution of the proper time and co-moving spatial coordinates relative to the coordinates of a particular foliation). The nine coordinate components $E_a{}^i$ describe projections of the spatial tetrads tangent to the constant-time hypersurface $\Sigma_t$. 
Note that the spatial triad vectors have zero time component, since we have chosen our tetrad frame gauge to be hypersurface-orthogonal. (For arbitrary tetrad frame gauge choices, this is not the case.) 

The $E_a{}^i$ are dynamical variables determined by the evolution and constraint equations
 \begin{eqnarray}
 \label{eq-E-ai}
N^{-1}\partial_t E_a{}^i &=& - K_a{}^c E_c{}^i 
,\\
\label{const-E-ai}
\epsilon_c{}^{ab}E_a{}^i \partial_i E_b{}^j &=& N^d{}_c E_d{}^j - N^d{}_d E_c{}^j,
\end{eqnarray}
 which are derived from applying the commutators of the basis vectors to the spatial coordinates $\{x^i\}$.
 
The lapse function $N$ and the shift vector $N^i$ are gauge variables that we can freely specify. A natural coordinate gauge choice when studying the evolution of cosmological spacetimes is to have co-moving coordinates, such that the $x^i$ are constant along the congruence and 
\begin{equation}
N^i \equiv 0.  
 \end{equation}
 
For the lapse function, we will impose the condition that hypersurfaces $\Sigma_t$ of constant time be constant mean curvature (CMC) hypersurfaces, {\it i.e.}, the trace of the extrinsic curvature $K_{ab}$ is spatially uniform for each $\Sigma_t$,
\begin{equation}
\label{CMC}
\Theta^{-1} \equiv - {\textstyle \frac13}K_a{}^a \big|_{\Sigma_t} = {\rm const} > 0 .
\end{equation}

Imposing CMC slicing, the trace of Eq.~\eqref{eq-K-ab} reduces to a linear, elliptic equation for the lapse $N$,
\begin{equation}
\label{lapse-eq}
\Big( - \big(D_a - 2A_a\big)D^a + \Sigma_{ab}\Sigma^{ab} +  3\Theta^{-2} + W^2-V(\phi)\Big)N =0,
\end{equation}
where 
\begin{equation}
\label{sigma-ab-def}
\Sigma_{ab}\equiv K_{ab} - {\textstyle \frac13}K_c{}^c\delta_{ab} 
\end{equation} 
is the trace-free part of the extrinsic curvature. Few works that rigorously prove well-posedness treat elliptic gauge conditions, and we are not aware of any for the particular tetrad formulation we use; however, see Ref.~\cite{Andersson:2001kw} for a proof in a closely related coordinate based formulation. 
We note, though, that if a system of partial differential equations is not well posed it would 
be challenging, if not impossible, to obtain stable, convergent numerical solutions
with any discretization scheme. That our code is stable and convergent
can thus be viewed as `empirical evidence' that the underlying equations are well posed.

Note that the (elliptic) equation for the (tetrad) lapse also determines the tetrad gauge source function $b_a$, which denotes the acceleration of the tetrad congruence worldlines. This can be seen, {\it e.g.}, by computing the commutator $[e_0, e_a]$ as applied to the time coordinate $x^0$,
\begin{equation}
b_a= N^{-1} E_a{}^i\partial_i N. 
\end{equation}

For the time coordinate $t$, we choose a particular (re-)scaling,
\begin{equation}
\label{timechoice}
{e^t} = 3 \Theta,
\end{equation}
that is consistent with the CMC slicing condition. If Eq.~\eqref{CMC} is satisfied, the inverse trace of the extrinsic curvature (here and throughout denoted by $\Theta$) is the well-known Hubble radius (as measured by the proper time coordinate $\tau$), {\it i.e.}, 
\begin{equation}
{\textstyle \frac13}e^{-t} =  \frac{d \ln a(\tau)}{d\tau}.
\end{equation}

During contraction, the Hubble radius decreases. Accordingly, the time coordinate $t\leq0$, running from small negative to large negative values. Yet, due to CMC slicing, all curvature variables remain finite and non-zero for any evolution of finite duration. This is a necessary condition to stably evolve contracting phases that last several hundreds of $e$-folds, which is required to study the robustness of slow contraction.

In addition, for a sufficiently long evolution, we must ensure that no two dynamical variables grow (or shrink) at significantly different rates to avoid a {\it stiffness problem}. Since in the case of slow contraction, some spatial metric variables, such as the spatially averaged scale factor, decrease at a significantly lower rate than some curvature variables, such as the Hubble radius, we eliminate the latter by introducing dimensionless Hubble-normalized variables,
\begin{eqnarray}
N &\to& {\cal N} \equiv N/\Theta, \\ 
\Big\{ E_a{}^i, \Sigma _{ab}, A_a , n_{ab}, W, S_a, V \Big\} &\to& \Big\{ {\bar E}_a{}^i, {\bar \Sigma} _{ab}, {\bar A}_a , {\bar n}_{ab}, {\bar W}, {\bar S}_a, {\bar V} \Big\},
\end{eqnarray}
where ${\cal N}$ is the Hubble-normalized lapse function; bar denotes multiplication by the Hubble radius $\Theta$; and 
\begin{equation}
n_{ab}\equiv N_{(ab)} 
\end{equation} 
is the symmetric part of $N_{ab}$.
Substituting into Eq.~\eqref{lapse-eq}, yields an elliptic equation for the Hubble-normalized lapse ${\cal N}$
\begin{eqnarray}
\label{Neqn}
&-& \bar{E}^a{}_i \partial^i \left(\bar{E}_a{}^j \partial _j {\cal N}\right) + 2 
\bar{A}^a \bar{E}_a{}^i\partial _i {\cal N} + {\cal N} \left(3 + \bar{\Sigma} _{a b}\bar{\Sigma}^{a b} + \bar{W}^2   - \bar{V} \right) = 
3 \,.
\end{eqnarray}

Imposing CMC slicing as defined in Eqs.~(\ref{CMC}, \ref{timechoice}) and using Hubble-normalized variables in Eqs.~(\ref{eq-K-ab}-\ref{eq-N-ab}, \ref{eq-phi}-\ref{eq-S-a}, and \ref{eq-E-ai}), the gravitational quantities ${\bar E}_a{}^i,  {\bar \Sigma}_{ab},  {\bar n}_{ab}$, and ${\bar A}_a$ as well as the  scalar field matter variables $\phi, \bar{W}, \bar{S}_a$ satisfy the hyperbolic evolution equations
\begin{eqnarray}
\label{eq-E-ai-Hn}
\partial_t \bar{E}_a{}^i &=& - \Big({\cal N} - 1 \Big) \bar{E}_a{}^i - {\cal N} \,\bar{\Sigma}_a{}^b \bar{E}_b{}^i 
,\\
\label{eq-sigma-ab}
\partial _t \bar{\Sigma}_{ab} &=& - \Big( 3 {\cal N} - 1 \Big) \bar{\Sigma}_{ab}
- {\cal N} \Big( 2 \bar{n}_{\langle a}{}^c\, \bar{n}_{b \rangle c}
- \bar{n}^c{}_c \bar{n}_{\langle ab \rangle} 
- \bar{S}_{\langle a} \bar{S}_{b \rangle} \Big)
+ \bar{E}_{\langle a}{}^i\partial _i \Big(\bar{E}_{b \rangle}{}^i\partial _i {\cal N}\Big) 
\\
&-& {\cal N} \left( \bar{E}_{\langle a}{}^i \partial_i \bar{A}_{b \rangle}
-  \epsilon^{cd}{}_{(a} \Big( \bar{E}_c{}^i\partial_i \bar{n}_{b)d} - 2 \bar{A}_c \bar{n}_{b )d} \Big) 
  \right)
+  \epsilon^{c d}{}_{(a} \bar{n}_{b ) d} \bar{E}_c{}^i \partial_i{\cal N}
+ \bar{A}_{\langle a} \bar{E}_{b \rangle}{}^i \partial_i {\cal N} 
\nonumber
,\\
\label{eq-n-ab}
\partial _t \bar{n}_{ab} &=& - \Big({\cal N} - 1 \Big) \bar{n}_{ab} 
+ {\cal N} \Big( 2  \bar{n}_{(a}{}^c \bar{\Sigma}_{b)c}
-\epsilon^{cd}{}_{( a} \bar{E}_c{}^i \partial _i \bar{\Sigma} _{b ) d} \Big)   
- \epsilon^{cd}{}_{( a} \bar{\Sigma}_{b) d} \bar{E}_c{}^i \partial _i {\cal N} 
,\\
\label{eq-A-a}
\partial _t \bar{A}_a &=& - \Big( {\cal N} - 1 \Big)\bar{A}_a 
- {\cal N} \Big( \bar{\Sigma} _a{}^b \bar{A}_b - {\textstyle \frac12} \bar{E}_b{}^i\partial _i \bar{\Sigma} _a{}^b \Big)  
- \bar{E}_a{}^i \partial _i{\cal N} 
+ {\textstyle \frac12} \bar{\Sigma} _a{}^b \bar{E}_b{}^i \partial _i {\cal N} 
,\\
\label{eq-phi-Hn}
\partial_t \phi &=& {\cal N} \,\bar{W}
,\\
\label{eq-w-Hn}
\partial_t \bar{W} &=& - \Big(  3 {\cal N} -1 \Big) \bar{W} 
- {\cal N} \Big(\bar{V}_{,\phi}  + 2 \bar{A}^a \bar{S}_a -  \bar{E}_a{}^i \partial_i \bar{S}^a  \Big)
+ \bar{S}^a \bar{E}_a{}^i\partial _i {\cal N}
,\\
\label{eq-S-a-Hn}
\label{eq-barS-Hn}
\partial_t \bar{S}_a &=& - \Big(  {\cal N} - 1 \Big) \bar{S}_a 
- {\cal N}\Big( \bar{\Sigma}_a{}^b \bar{S}_b - \bar{E}_a{}^i\partial _i \bar{W} \Big)
+  \bar{W} \bar{E}_a{}^i \partial_i {\cal N} 
.
\end{eqnarray}
(Here angle brackets denote traceless symmetrization defined as $X_{\langle ab \rangle} \equiv X_{(ab)} - {\textstyle \frac13}X_c{}^c\delta_{ab}$.)
The system of Eqs.~(\ref{Neqn}-\ref{eq-barS-Hn}) will serve as our numerical scheme. Notably, this system is well-posed, as was shown, {\it e.g.}, in \cite{Bardeen:2011ip}, and, by construction, it admits stable numerical evolution that involves several hundreds of $e$-folds of contraction.

In addition, the same variables satisfy the constraint equations 
\begin{eqnarray}
\label{constraintG}
{\cal C}_{\rm G} &\equiv& 3 + 2 \bar{E}_a{}^i \partial _a \bar{A}^a 
- 3 \bar{A}^a \bar{A}_a
- {\textstyle \frac12 } \bar{n}^{ab} \bar{n}_{ab}
+ {\textstyle \frac14 } ( \bar{n}^c{}_c)^2 
- {\textstyle \frac12 } \bar{\Sigma}^{ab} \bar{\Sigma}_{ab}
\\
&-& {\textstyle \frac12 } \bar{W}^2 -  {\textstyle \frac12 } \bar{S}^a \bar{S}_a -   {\bar V} = 0
\nonumber
\,,\\
\label{constraintC}
( {\cal C}_{\rm C} )_a & \equiv & \bar{E}_b{}^i \partial _i {\bar \Sigma} _a{}^b
 - 3 {\bar \Sigma} _a{}^b \bar{A}_b - \epsilon _a{}^{b c} \bar{n}_b{}^d \bar{\Sigma}_{cd} - {\bar W} {\bar S}_a = 0
\,,\\
\label{constraintJ}
( {\cal C}_{\rm J} )_a &\equiv& \bar{E}_b{}^i \partial _i \bar{n}^b{}_a + \epsilon^{bc}{}_a \bar{E}_b{}^i\partial _i \bar{A}_c - 2 \bar{A}_b \bar{n}^b{}_a =0
\,,\\
\label{constraintS-phi}
( {\cal C}_{\rm S})_a &\equiv& {\bar S}_a - {\bar E}_a{}^i\partial _i\phi = 0
\,,\\
\label{constraintCOM}
( {\cal C}_{\rm E} )_a^i &\equiv& \epsilon^{bc}{}_a
\Big( \bar{E}_b{}^j \partial_j \bar{E}_c{}^i - \bar{A}_b \bar{E}_c{}^i \Big) - \bar{n}_a{}^d \bar{E}_d{}^i = 0.
\end{eqnarray}
The constraints are the Hubble-normalized version of Eqs.~(\ref{H-const}-\ref{N-ab-const}, \ref{const-phi}, \ref{const-E-ai}), again imposing CMC slicing conditions as in Eqs.~(\ref{CMC}, \ref{timechoice}). 
(The subscripts $G, C$ and $J$ stand for Gauss, Codazzi, and Jacobi, respectively, referring to the commonly used terminology.)
As detailed in the following two sections, we shall use the constraint equations to specify the initial conditions and to ensure constraint damping and numerical convergence.

\section{Initial conditions}
\label{sec:initialdata}

In testing the robustness of slow contraction, the set of initial conditions under study plays a key role. Analytic-perturbative analyses of smoothing mechanisms can only establish `classical smoothing,' {\it i.e.} stability of the attractor FRW solution to small inhomogeneities and anisotropies \cite{Cook:2020oaj}. To establish `robustness,' the evolution must be studied under a wide set of initial conditions  including those that are far outside the perturbative regime of the FRW attractor solution. As we shall describe in this section, our scheme enables the variation of all freely specifiable variables, such as the initial shear and spatial curvature contributions, $\{\bar{\Sigma}_{ab}, \bar{A}_a\}$, as well as the initial field and velocity distributions of the scalar, $\{\phi, \bar{W}\}$. 

\subsection{Geometric variables}

To specify the initial conditions, we first choose a particular time $t_0$.  With the tetrad and coordinate gauge conditions remaining the same as specified above for the evolution scheme, the $t_0$-hypersurface has constant mean curvature $\Theta^{-1}_0$, the value of which we can freely choose, and zero shift. We note that, using Hubble-normalized variables in our evolution scheme,  the natural units are set by this initial Hubble radius $\Theta_0$ rather than the Planck scale.

Next, we must fix the variables 
\begin{equation}
\big\{\bar{E}_a{}^i, \bar{n}_{ab}, \bar{A}_a, \bar{\Sigma}_{ab} \big\}
\end{equation}
that describe the geometry of the $t_0$-hypersurface.
Not all of these variables are freely specifiable, though, as they must satisfy the constraint equations~(\ref{constraintG}-\ref{constraintCOM}). (Notably, the evolution equations~(\ref{eq-E-ai-Hn}-\ref{eq-S-a-Hn}) propagate the constraints, {\it i.e.,} ensure that the constraints are satisfied at later times provided they are satisfied on the initial time slice.)

Adapting the York method \cite{York:1971hw} commonly used in numerical relativity computations,  we choose  the initial metric to be conformally flat,
\begin{equation}
g_{ij} \equiv \psi^4(x, t_0) \delta_{ij},
\end{equation}
where $\psi$ is the conformal factor. The conformal factor is not a free function but fixed by the Hamiltonian (or Gauss) constraint~\eqref{constraintG}, as we will detail below.

With Eq.~\eqref{g-mu-nu-rel}, this choice for $g_{ij}$ simultaneously fixes the coordinate components of the spatial triad:
\begin{equation}
\label{init-E-ab}
{\bar E}_a{}^i =  \psi ^{-2} \Theta^{-1}_0 \delta _a{}^i.
\end{equation}
Substituting into the spatial commutator, 
\begin{equation}
\label{comm-sp}
[\bar{e}_a, \bar{e}_b] \equiv   \bar{{\cal C}}_{abc} \bar{e}^c = \Big( 2 \bar{A}_{[a}\delta_{b]c} + \epsilon_{abd}\bar{n}^d{}_c\Big) \bar{e}^c
,
\end{equation}
as defined in Eq.~\eqref{comm}, we find the components of the intrinsic curvature tensor,
\begin{eqnarray}
\label{init-n-ab}
{\bar n}_{ab} &=& 0
,\\
\label{init-A-a}
{\bar A}_a &=& - 2 \psi^{-1} \bar{E}_a{}^i \partial _i \psi ;
\end{eqnarray}
and the constraint equations \eqref{constraintJ} and \eqref{constraintCOM} are trivially satisfied by this combination of $\bar{E}_a{}^i, \bar{n}_{ab}, \bar{A}_a$.
  
We stress that, in general, ${\bar A}_a$ is non-zero, reflecting the fact that the anti-symmetric part of the intrinsic curvature tensor does not transform trivially under conformal rescaling, as pointed out in Ref.~\cite{Bardeen:2011ip}. That means, most especially, assuming the initial metric to be conformally flat does not imply uniform spatial curvature on the initial slice and, hence, does not impose a real restriction on the initial data set. 

Having set the geometric variables $\{\bar{E}_a{}^i, \bar{n}_{ab}, \bar{A}_a\}$ through specifying the spatial metric for the initial slice, it remains to determine the components of the Hubble-normalized shear tensor $\bar{\Sigma}_{ab}$, as defined in Eq.~\eqref{sigma-ab-def}, to close the set of variables describing the geometry of the initial $t_0$-hypersurface. 
 
It is a particular advantage of the conformal rescaling as suggested by the York method that the constraint equations significantly simplify. For this reason, we will determine the initial shear contribution $\bar{\Sigma}_{ab}(x,t_0)$ by first specifying its conformally rescaled counterpart 
\begin{equation}
\label{confZ}
Z_{ab}(x,t_0) \equiv \psi^6 \bar{\Sigma} _{ab}(x,t_0),
\end{equation}
using the momentum constraint~\eqref{constraintC} as evaluated for $Z_{ab}(x, t_0)$, 
\begin{equation}
\label{divZ}
\bar{E}^a{}_i(x, t_0) \,\partial^i Z_{ab} (x, t_0) = Q(x, t_0) \bar{E}_b{}^i (x, t_0) \partial_i \phi(x, t_0).
\end{equation}
Here, $Q(x, t_0)$ is the conformally rescaled scalar field kinetic energy defined by
\begin{eqnarray}
\label{def-Q}
Q(x,t_0) &\equiv & \psi^6(x,t_0)\bar{W}(x,t_0).
\end{eqnarray}
Then, solving the Hamiltonian constraint \eqref{constraintG} for the conformal factor $\psi(x,t_0)$ yields $\bar{\Sigma}_{ab}(x,t_0)$.

It is immediately obvious that we can follow two strategies in specifying $Z_{ab}(x,t_0)$: either we freely specify the initial shear contribution $Z_{ab}(x, t_0)$ that then fixes parts of the initial field and velocity distribution $\{\phi(x, t_0), Q(x, t_0)\}$, or we freely specify only parts of the initial shear contribution $Z_{ab}(x, t_0)$ so we can freely choose $\{\phi(x, t_0), Q(x, t_0)\}$ that together fix the rest of $Z_{ab}(x, t_0)$ using the momentum constraint. In this analysis, we have chosen the latter option.

By definition, the vacuum contribution $Z_{ab}^0(x, t_0)$ of the conformally rescaled shear tensor, {\it i.e.}, the divergence-free part of $Z_{ab}(x, t_0)$, is independent of any matter source. Accordingly, we will freely specify only $Z_{ab}^0(x, t_0)$ and constrain the rest of $Z_{ab}(x, t_0)$ by the choice of the initial scalar field and velocity distribution $\{\phi(x, t_0), Q(x, t_0)\}$.

For the numerical simulations, we use periodic boundary conditions $0 \le x \le 2 \pi$ with $0$ and $2 \pi$ identified. In addition, we restrict ourselves to deviations from homogeneity
along a single spatial direction $x$ so that the spacetimes have two Killing fields. 
Since the variables depend only on $x$ and since $x$ is periodically identified, we specify their spatial variation as a sum of Fourier modes with period $2\pi$.  

A simple yet generic divergence-free and trace-free  choice for $Z_{ab}^0(x, t_0)$ is given by
%
%
\begin{equation}
\label{zic}
Z_{ab}^0(x, t_0) = \left(
\begin{array}{ccc}
{b_2} & \xi & 0\\
\xi & a_1\cos x + {b_1} & a_2\cos x\\
0 & a_2\cos x & -{b_1}-{b_2}-a_1\cos x
\end{array}\right),
\end{equation}
where $\xi ,\, a_1, \, a_2,\  b_1$ and $b_2$ are constants. Note that since $\bar{\Sigma} _{ab}$ must be trace-free, so must $Z^0_{ab}$. The rest of the initial shear distribution, $Z_{ab} - Z^0_{ab}$ is obtained by solving the momentum constraint~\eqref{divZ}, which turns into an algebraic relation for the Fourier coefficients of this non-zero 
divergence piece of $Z_{ab}$ upon specifying the initial scalar field matter variables.

\subsection{Scalar-field matter variables}

In setting the initial velocity and field distribution, we freely specify the Fourier coefficients of $Q(x, t_0)$ and $\phi(x, t_0)$ via
\begin{eqnarray}
\label{phiic}
\phi(x, t_0) &=& f_0 \cos \big(m_0 x + d_0 \big)
,\\
\label{Qic}
Q(x,t_0) &=& \Theta \Big(f_1  \cos \big(m_1 x + d_1 \big) + Q_0 \Big).
\end{eqnarray}
where $f_0, m_0, d_0, f_1, m_1, d_1$, and $Q_0$ are constants.

To compute the initial value of the (Hubble-normalized) scalar field gradient $\bar{S}_a(x, t_0)$, we substitute $\phi(x, t_0) $ into the constraint equation \eqref{constraintS-phi}.
 
 \subsection{Conformal factor}
 
Finally, imposing the Hamiltonian constraint~\eqref{constraintG}, our {\it ansatz} for the initial set of geometric and matter variables yields an elliptic equation for the conformal factor $\psi$:
\begin{eqnarray} \label{conformal}
\partial^i \partial _i \psi &=& {\textstyle \frac14 } \left( 3 \Theta^{-2} - V  \right) \psi^5 - {\textstyle \frac18 } \left( \partial^i \phi \partial _i \phi \right)\psi
- {\textstyle \frac18 } \left( Q^2  + Z^{ik} Z_{ik} \right) \Theta^2 \psi^{-7}.
\end{eqnarray} 
We numerically solve these equations in the following section.

\section{Numerical Tests of Robustness}
\label{sec_results}

Using the numerical general relativity scheme described in Secs.~\ref{sec_methods} and~\ref{sec:initialdata}, we have conducted extensive series of studies to gauge the robustness of 
 slow contraction in smoothing and flattening the universe beginning from 
 a broad spectrum of highly non-perturbative, inhomogeneous initial matter, spatial curvature and shear 
 distributions.  To date, there has been no analogous test of robustness for an expanding cosmology, including inflation; see discussion in Ref.~\cite{Cook:2020oaj}.
 
As described in Sec.~\ref{sec:initialdata}, we solve the full $3+1$-dimensional Einstein-scalar field equations beginning from arbitrary combinations of shear ($\Omega_s$), spatial curvature ($\Omega_k$) and matter ({\it i.e.}, scalar field) density ($\Omega_m$), where 
\begin{eqnarray}
\Omega_s & \equiv & {\textstyle\frac16} \bar{\Sigma}^{ab} \bar{\Sigma}_{ab}  ,
\\
\Omega_m& \equiv & {\textstyle\frac16} \bar{W}^2 + {\textstyle\frac16} \bar{S}^a \bar{S}_a + {\textstyle\frac13} \bar{V} ,
\\
\Omega_k  & \equiv & - {\textstyle\frac23} \bar{E}_a{}^i\partial_{i} \bar{A}^a +  \bar{A}^a \bar{A}_a + {\textstyle\frac16} \bar{n}^{ab} \bar{n}_{ab} - {\textstyle\frac{1}{12}}(\bar{n}^c{}_c)^2,
\end{eqnarray}
and $\sum_i \Omega_i =1$.
For simplicity, all deviations from homogeneity are along a single spatial direction $x$, as in Ref.~\cite{Cook:2020oaj}.  

The scalar field potential energy takes the form
\begin{equation} 
\label{pot}
V(\phi) = V_0\, {\rm exp} \big(- \sqrt{2 \varepsilon}\, \phi \big)
\end{equation}
where $V_0 <0$ and $\varepsilon \gg 3$.   The dynamical $\Omega_m=1$ FRW attractor solution corresponds to the effective equation-of-state  $\varepsilon_{\rm eff} \rightarrow \varepsilon$.
The initial spatial inhomogeneities in the matter density are set by $f_1$ in the expression for $Q(x,t)$ in Eq.~(\ref{Qic}).   More precisely, $Q(x,0)$ is the initial velocity distribution, and  $f_1$ is the magnitude of a velocity fluctuation mode with wavenumber $m_1$ about the mean initial scalar field velocity $Q_0$ at $t=0$.  The sign convention is that {\it positive $Q(x,t)$ corresponds to rolling down the potential} (towards more negative values of  $V(\phi)$).   Spatial variations in the initial shear $Z_{ab}^0$ are set by the parameters $a_1$ and $a_2$ in Eq.~(\ref{zic}).
Once these parameters are set, the initial conditions for the spatial curvature, shear and matter density are completed by computing the conformal factor $\psi$ in Eq.~(\ref{conformal}) and using Eqs.~(\ref{confZ}) and~(\ref{def-Q}), as described in Sec.~\ref{sec:initialdata}.
  The studies consisted of a series of simulations in which each of these parameters that set the initial conditions was varied independently from zero to ${\cal O}(1)$.   

 Parameters that were found not to affect significantly the robustness tests were held fixed. For the purposes of the simulations, the coefficient of the potential in Eq.~(\ref{pot}) was set to $V_0=-0.1$ (in units of the initial $\Theta$); the initial scalar field was set to $\phi(x,t)=0$;
 the period and shift of the sinusoidal spatial variation of scalar field velocity $Q(x,t)$ in Eq.~(\ref{Qic}) were set to $m_1=1$; $d_1=0$; and the remaining constant coefficients in $Z_{ab}^0$ in Eq.~(\ref{zic}) were set to $\xi=0.01$,   $b_1= -0.15$ and $b_2=-1.8$.

The result of the numerical studies can be summarized in a series of `phase diagrams'  indicating the final state as a function of the initial mean scalar field velocity $Q_0$ and potential energy density parameter $\varepsilon$.  The different phase diagrams correspond to different types of initial conditions as described in the text below.
The phase diagrams were made by first making runs for a  coarse sampling of  $Q_0$ and $\varepsilon$ and characterizing the final states at the end of a representative run lasting $\mathfrak{N}_H=200$ e-folds, where $\mathfrak{N}_H$ is equal to the number of e-folds of contraction of the Hubble radius or, equivalently, $\Theta = |H|^{-1}$.  Then further runs were made, holding $\varepsilon$ fixed and performing a bisection search in $Q_0$ to identify the boundaries between different end states to 1 decimal place precision. By comparing results at different resolutions, we conclude that the dominant numerical error in our results arises from the precision used in the bisection search. We note also that smoothing occurs very slowly for the lowest value of $\varepsilon$ considered, $\varepsilon=4.5$, and longer lasting runs suggest that the simulations that have not smoothed, or only partially smoothed by $\mathfrak{N}_H=200$, will continue to smooth further. Nevertheless, the diagrams below are based on the end state at $\mathfrak{N}_H=200$  for consistency.  For  $\varepsilon \approx 4.5$, our results should be viewed as conservative upper bounds for smoothing.

\begin{figure*}[t!]
\begin{center}
\includegraphics[width=3.00in,angle=-0]{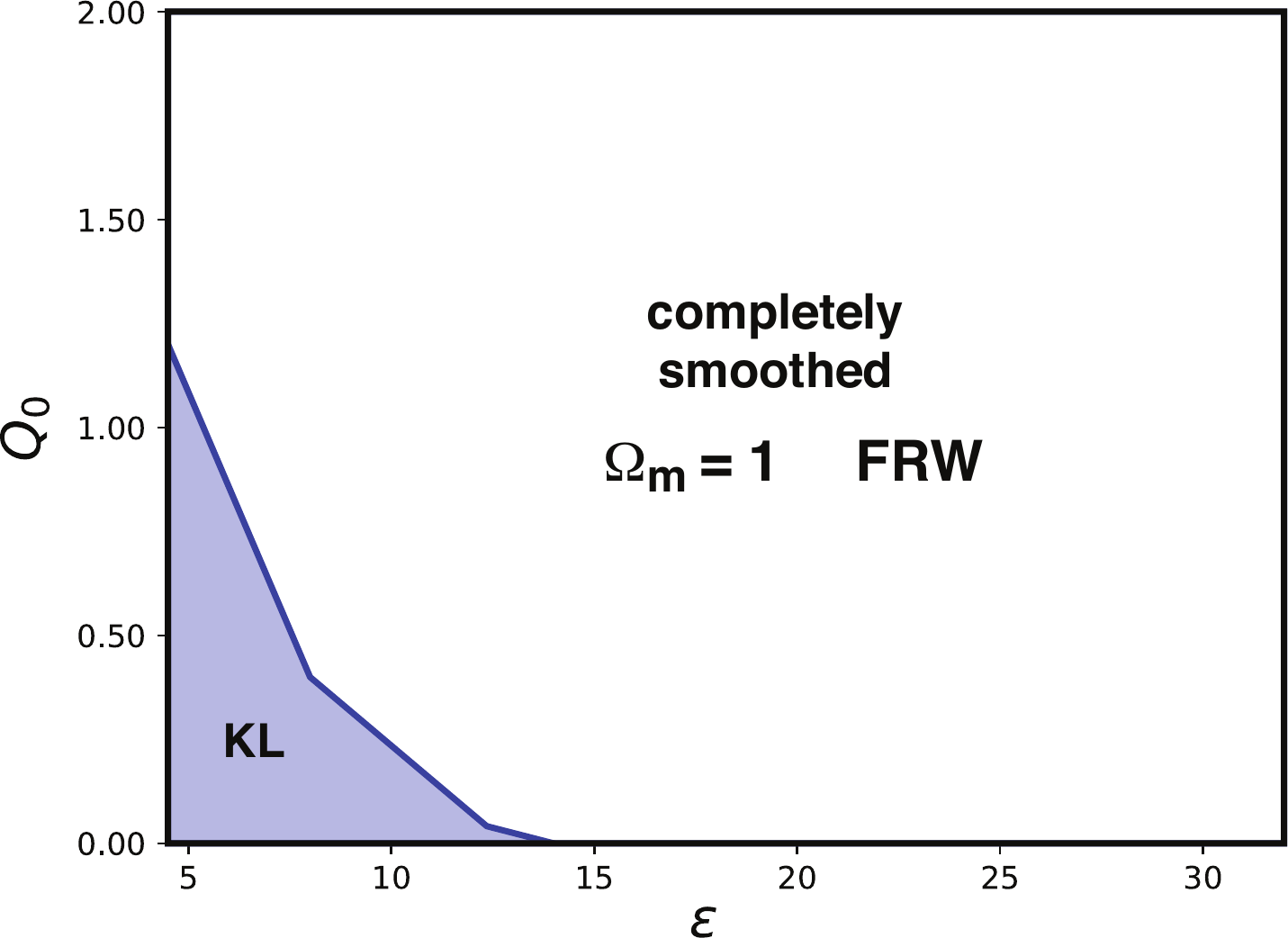}
\end{center}
\caption{Phase Diagram I shows the two possible final states ($\Omega_m=1$ FRW and KL) reached for the case of homogeneous initial conditions as a function of the mean initial scalar field velocity $Q_0$ and the scalar field potential parameter $\varepsilon$. The curve divides the diagram into two regimes.     
\label{fig:1}}
\end{figure*}    

\vspace{.1in}
\noindent
{\sc Phase Diagram I.} We first consider the case of fully homogeneous initial conditions: $f_1=a_1=a_2=0$.    In this case, the simulations converge to one of two homogeneous fixed-point outcomes:  
\begin{itemize}[leftmargin=*]
\item an $\Omega_m=1$ FRW universe with $\Omega_k=\Omega_s=0$, in which the matter density consists of a uniform combination of scalar field kinetic and potential energy (no scalar field gradient energy density) that has converged to the dynamical attractor solution for all $x$:  namely, 
$\varepsilon_{\rm eff} \rightarrow \frac{3}{2} \dot{\phi}^2/ (\frac{1}{2} \dot{\phi}^2+V) \rightarrow \varepsilon \gg 3$; or, 
\item a ``Kasner-like'' (KL) universe that is homogeneous, spatially flat ($\Omega_k=0$) and comprised of some uniform mixture of $\Omega_m$ and $\Omega_s$ such that $\Omega_m+\Omega_s=1$;  the matter density consists purely of scalar field kinetic energy density (with zero gradient or potential energy density) corresponding to $\varepsilon_{\rm eff}\rightarrow 3$; the ratio of $\Omega_m$ to $\Omega_s$ at the fixed point depends on the initial values of the $\Omega_i$.
\end{itemize} 
%
\begin{figure*}[tbh!]
\begin{center}
\includegraphics[width=3.675in,angle=-0]{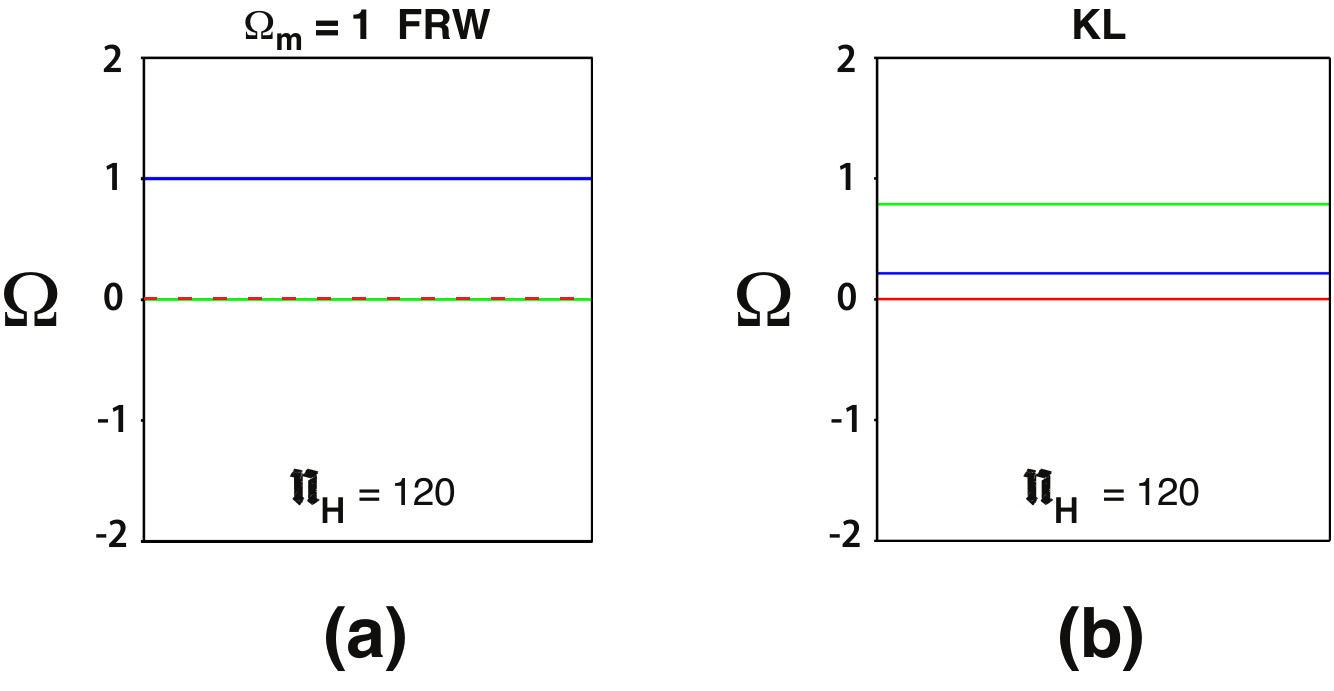}
\end{center}
\caption{Plots of the final states corresponding to the two phases shown in Phase Diagram I:  (a) the dynamical attractor 
 $\Omega_m=1$ FRW state; and (b) the homogeneous Kasner-like (KL) state with non-zero $\Omega_s$ and $\Omega_m$.  both final states have $\Omega_k=0$.   Each plot shows 
normalized energy density 
in matter $\Omega_m$
(blue), curvature $\Omega_k$ (red) and
shear $\Omega_s$ (green) as a function of  
$0\le x \le 2 \pi$.  $\mathfrak{N}_H$ is equal to the number of e-folds of contraction of $\Theta = |H|^{-1}$.
\label{fig:2}}
\end{figure*} 

The phase diagram in Fig.~\ref{fig:1} shows a curve that divides the plot into two regions.  (The curve is identical to the one marked $\Delta f_1=0$ curve in Fig.~3 of Ref.~\cite{Cook:2020oaj}.). All initial conditions above the curve converge to the $\Omega_m=1$ FRW fixed point and all initial conditions below converge to the KL fixed point.   Snapshots of the final distributions of the $\Omega_i$'s for the two types of fixed points are shown in Fig.~\ref{fig:2}.

Note that the curve falls below $Q_0=0$ for $\varepsilon \gtrsim 13$. 
As discussed in Ref.~\cite{Cook:2020oaj}, in cyclic and most bouncing cosmological models, the natural initial condition at the onset of a contracting phase is that the scalar field is either at rest or evolving down the potential, though with speeds that can vary with $x$.  This corresponds to mean initial scalar field velocity $Q_0\ge 0$ or `down the potential.'  Also, as shown in Ref.~\cite{Cook:2020oaj}, practical bouncing models require $\varepsilon \gtrsim 13$ in order that the smoothing and flattening (beginning from non-perturbative initial conditions) is completed rapidly enough that a nearly scale-invariant spectrum of density perturbations consistent with observations can be generated from quantum fluctuations before the bounce.  Consequently, the remainder of the discussion will be confined to  $Q_0\ge 0$ and, while we will comment on some interesting behaviors with $\varepsilon < 13$, the focus, as far as realistic applications to cyclic and bouncing cosmology are concerned, should be on the results when $\varepsilon \gtrsim 13$.  In Phase Diagram I, the fixed point in this entire region converges to the desired $\Omega_m=1$ FRW dynamical attractor.
\begin{figure*}[b!]
\begin{center}
\includegraphics[width=3.00in,angle=-0]{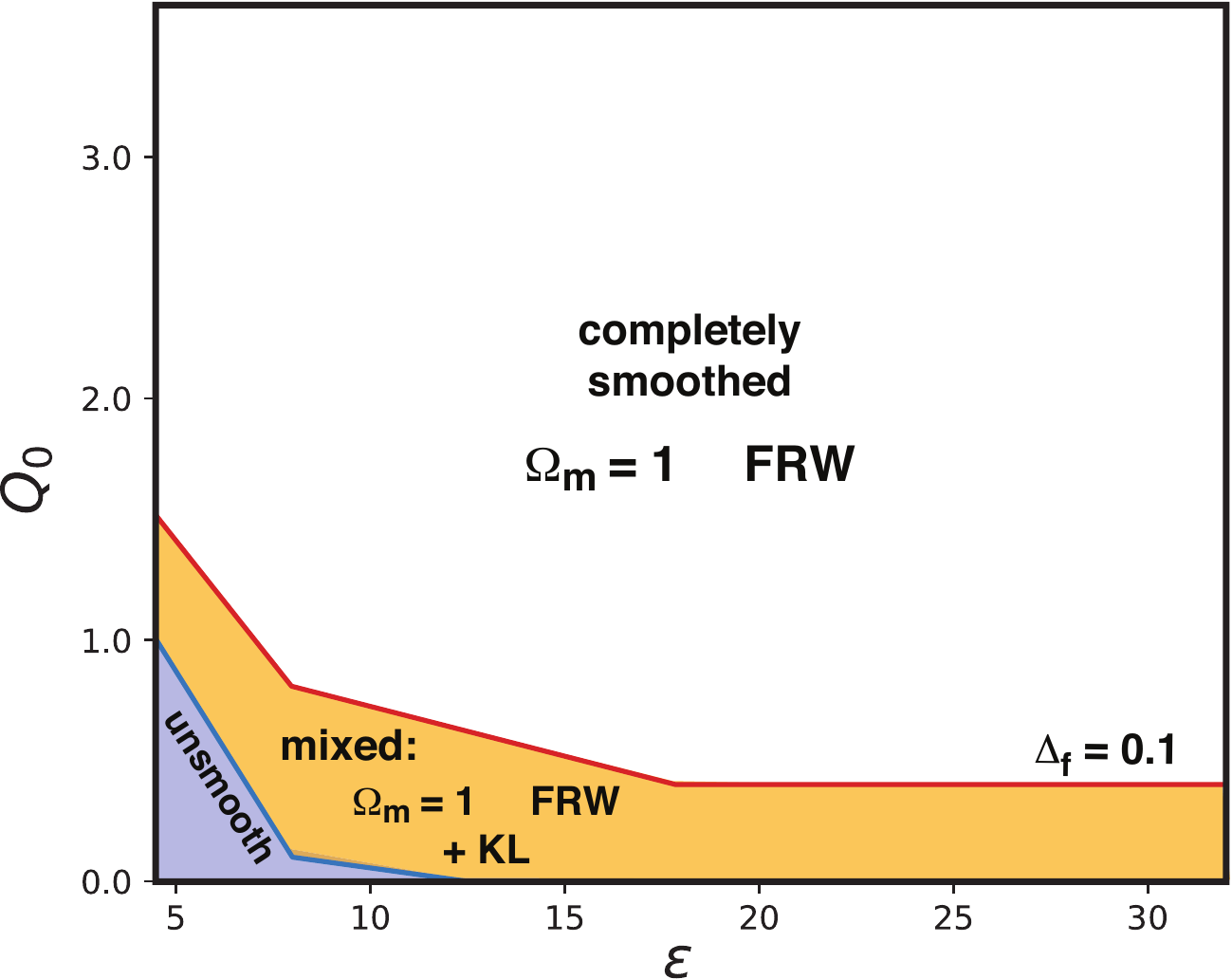}
\end{center}
\caption{Phase Diagram II shows the final states reached for the cases in which the initial scalar field velocity is highly inhomogeneous ($\Delta_f = 0.1$) but the $Z_{ab}^0$ is homogeneous.  Over most of the phase diagram, spacetime is completely smoothed ($\Omega_m=1$ FRW) or in a mixed state that is smoothed to an exponential degree (as measured by proper volume). 
\label{fig:3}}
\end{figure*} 

\vspace{0.1in}
\noindent
{\sc Phase Diagram II:} The second phase diagram (Fig.~\ref{fig:3}) corresponds to initial conditions in which $Z_{ab}^{0}$ is homogeneous ($a_1=a_2=0$) but the initial scalar field velocity distribution is not ($f_1 \ne 0$).  Here the division between phases depends on  $\Delta_f= f_1/Q_{attr}$, where $Q_{attr}$ is the value of $Q(x,t)$ for the dynamical attractor solution.  $\Delta_f={\cal O}(1)$ corresponds to large initial velocity fluctuations in which the initial velocity deviates far from the attractor as a function of $x$.  We show two bounding curves in the plot (which correspond to two of the examples shown in Fig.~3 of Ref.~\cite{Cook:2020oaj}). For the practical reasons described above, we restrict the plot to the regime relevant for cyclic and bouncing cosmology, $Q_0>0$.

\begin{figure}[bt!]
\begin{center}
\includegraphics[width=5.5in,angle=-0]{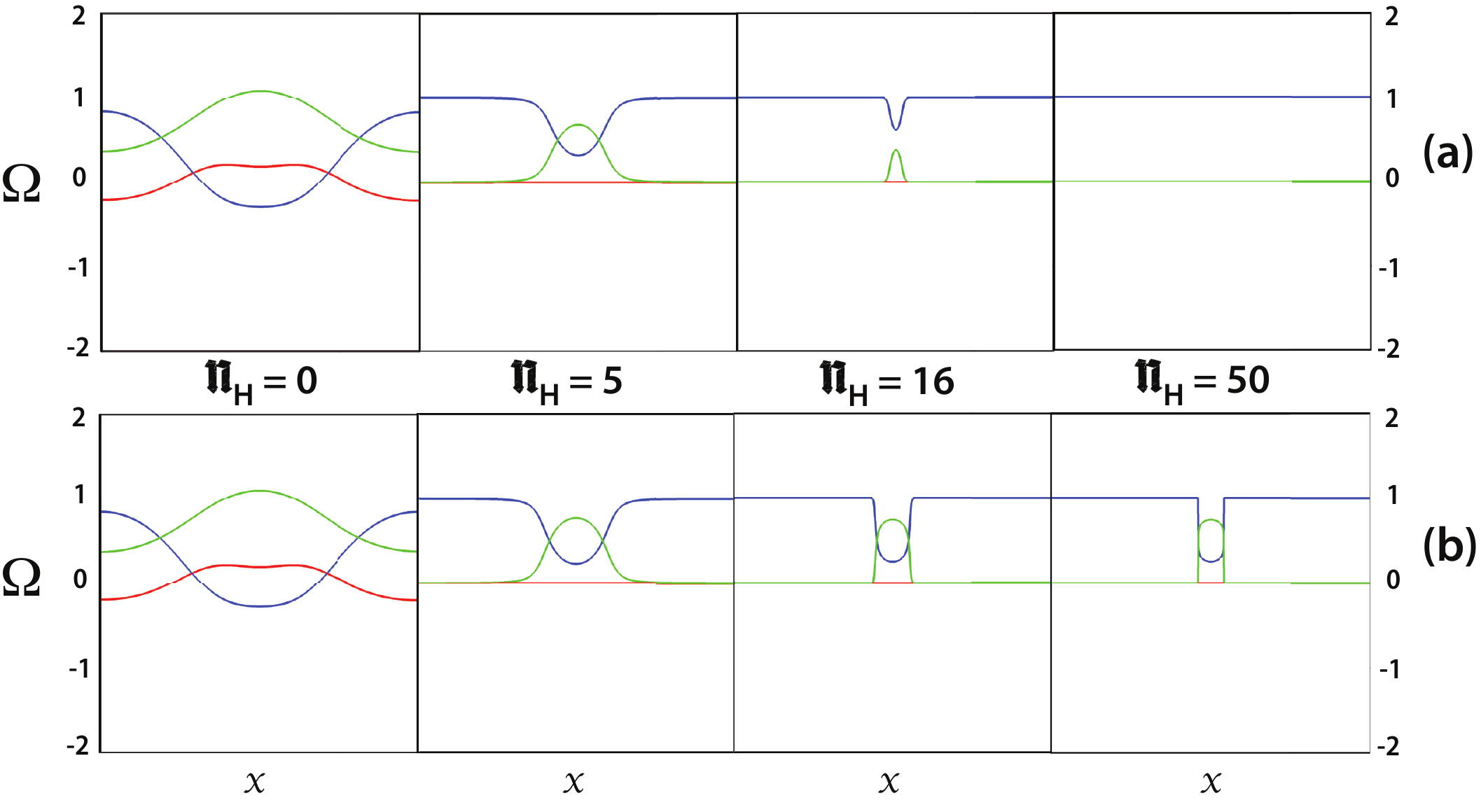}
\end{center}
\caption{Series of snapshots showing the evolution of the normalized energy density 
in matter $\Omega_m$
(blue), curvature $\Omega_k$ (red) and
shear $\Omega_s$ (green) 
for two cases with $\Delta_f = 0.1$ drawn from Phase Diagram II: (a) for $Q_0=0.8$, convergence to the  $\Omega_m=1$ FRW dynamical attractor with $\epsilon_{\rm eff} \rightarrow \epsilon =8$; and (b) for $Q_0=0.7$,  convergence to a mixed state in which space is divided into segments that are $\Omega_m=1$ FRW separated by segments that are KL but spatially varying.
 $\mathfrak{N}_H$ is equal to the number of e-folds of contraction of $\Theta = |H|^{-1}$.
\label{fig:4}}
\end{figure} 

\begin{figure*}[tb!]
\begin{center}
\includegraphics[width=4.50in,angle=-0]{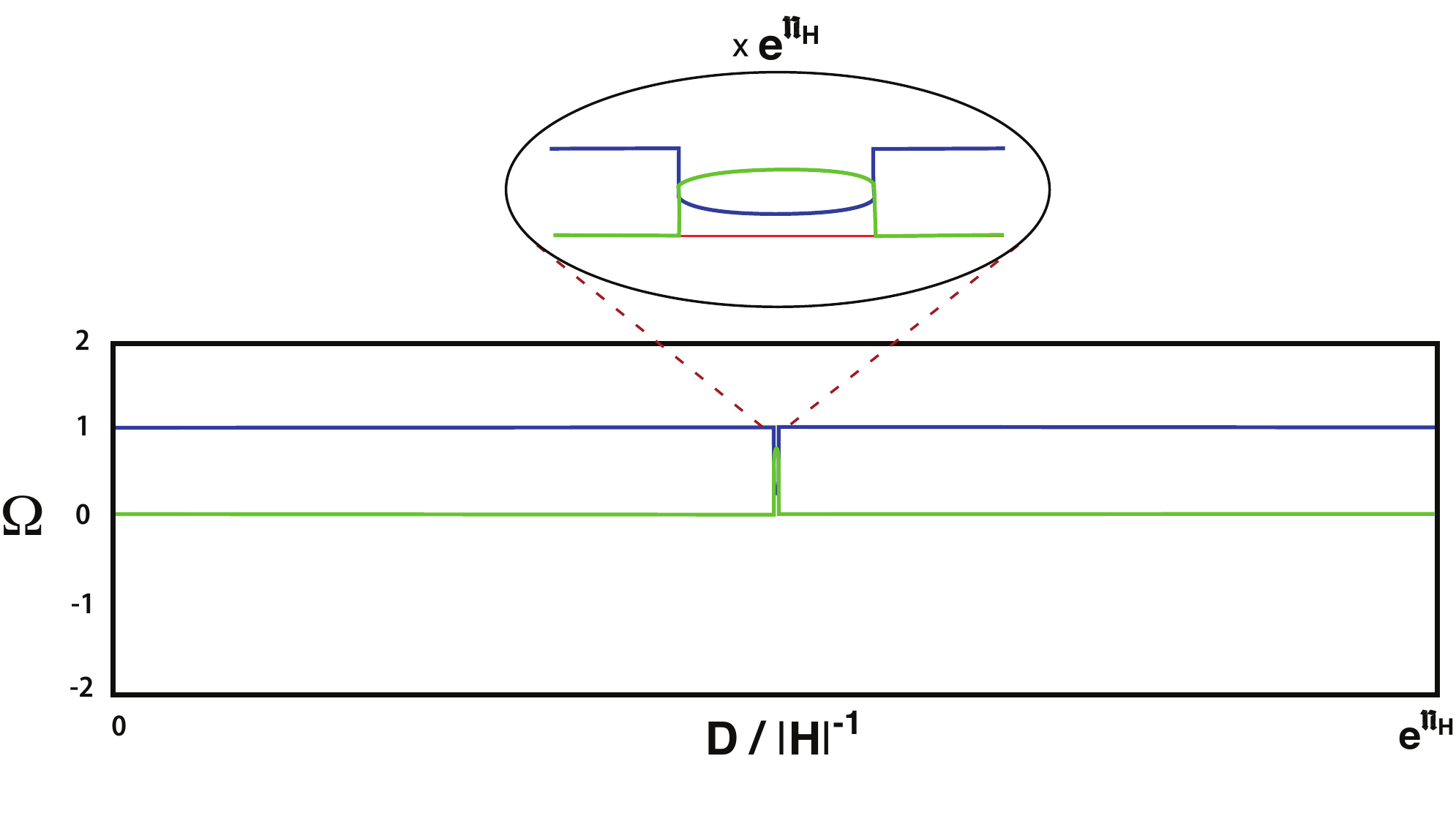}
\end{center}
\caption{Sketch of the final (mixed) state shown in Fig.~\ref{fig:4}(b) in physical distance coordinate $D$ compared to the Hubble radius $|H|^{-1}$ showing that the spatially varying KL segment is exponentially small compared to the homogeneous $\Omega_m=1$ FRW region that occupies most of the spacetime volume.
\label{fig:4half}}
\end{figure*}

In this case, initial conditions corresponding to points above the curve converge to the  $\Omega_m=1$ FRW dynamical attractor.   Initial conditions below the curve  evolve into ``mixed states'' in which the volume appears to be divided into segments that converge to the dynamical attractor  with $\varepsilon_{\rm eff} = \varepsilon \gg 3$ separated by segments that are `KL but spatially varying,'  as shown in Fig.~\ref{fig:4}.  The latter refers to segments in which $\Omega_k=0$  and $\varepsilon_{\rm eff} =3$ (as in case of homogeneous KL), but the ratio of $\Omega_m$ to $\Omega_s$ varies continuously across the CMC hypersurfaces.   Because the physical volumes corresponding to each segment of length $\Delta x$ contract as 
\begin{equation}
a^3(\tau) \Delta x = |\tau|^{3/\varepsilon_{\rm eff}}  \Delta x,
\end{equation}
where $\tau$ is the proper FRW time coordinate, the slowly contracting  $\Omega_m=1$ FRW segments with $\varepsilon_{\rm eff} \gg3$ become exponentially larger than the KL segments with $\varepsilon_{\rm eff}=3$ as contraction proceeds.  The situation is illustrated in Fig.~\ref{fig:4half} where the main plot is the physical distance divided by the Hubble radius; the spatially varying KL segment shown in the inset is exponentially small by comparison.
Similar remarks apply to all the examples of mixed states below.  ({\sc n.b.} This more intuitive argument agrees with the more formal proper volume analysis in Ref.~\cite{Garfinkle:2008ei}.)

In other words, except for the sliver of small $\epsilon$ and $Q_0$ marked `unsmooth,' all initial conditions in the phase diagram either converge  to the dynamical attractor solution for all $x$ or into mixed states in which an exponentially large fraction of space time converges to the dynamical attractor but there are also (cosmologically irrelevant) infinitesimal regions with spatially varying KL behavior.  

\begin{figure*}[b!]
\begin{center}
\includegraphics[width=5.00in]{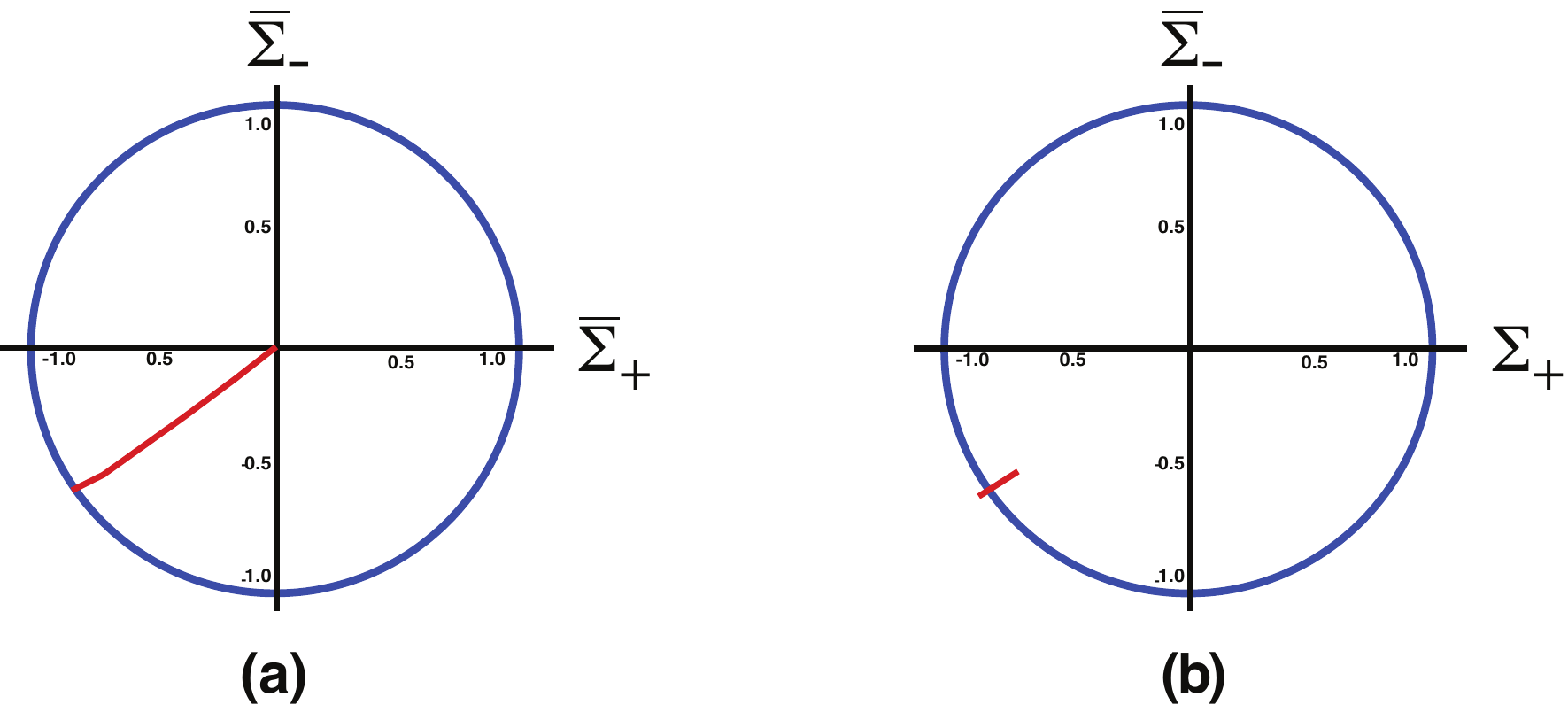}
\end{center}
\caption{The state space orbits comparing worldlines at
$x=\pi$ for the two models in Fig.~\ref{fig:4}: (a)  converges to  the  $\Omega_m=1$ FRW dynamical attractor, and (b) evolves into a mixed state that includes a segment corresponding to spatially varying KL.   The point $x=\pi$ lies in the spatially varying KL region in the second example; the corresponding orbit never reaches the center corresponding to the  $\Omega_m=1$ FRW dynamical attractor.
\label{fig:5}}
\end{figure*}
Fig.~\ref{fig:5} shows  state space orbit plots associated with two examples in Fig.~\ref{fig:4}.  The orbit plots enable the visualization of the evolution of the shear at a chosen point $x$ in the
$(\bar{\Sigma}_+ ,\bar{\Sigma}_-)$ plane, where
\begin{equation}\label{Sigmaplus}
\bar{\Sigma}_+ = \textstyle{\frac12}\Big(\bar{\Sigma}_{11} + \bar{\Sigma}_{22}\Big)
,\quad
\bar{\Sigma}_- = \textstyle{\frac{1}{2\sqrt{3}}}\Big(\bar{\Sigma}_{11} - \bar{\Sigma}_{22}\Big).
\end{equation}
The $\bar{\Sigma}_{\pm}$ are normalized so that the unit circle ($\bar{\Sigma}_+^2+\bar{\Sigma}_-^2 =1$) corresponds to the vacuum Kasner solution; trajectories that approach $\Omega_m=1$ FRW  converge to the center. 
See Refs.~\cite{Garfinkle:2008ei,Cook:2020oaj} for other examples.  In the case of Fig.~\ref{fig:5}, the orbit converges to the center ($\Omega_m=1$  FRW) for the case in Fig.~\ref{fig:4}a for all $x$; for the mixed state case in  Fig.~\ref{fig:4}b, the point $x = \pi$ lies in the spatially varying KL region between the Kasner circle and the origin. 

\begin{figure*}[t!]
\begin{center}
\includegraphics[width=6.00in,angle=-0]{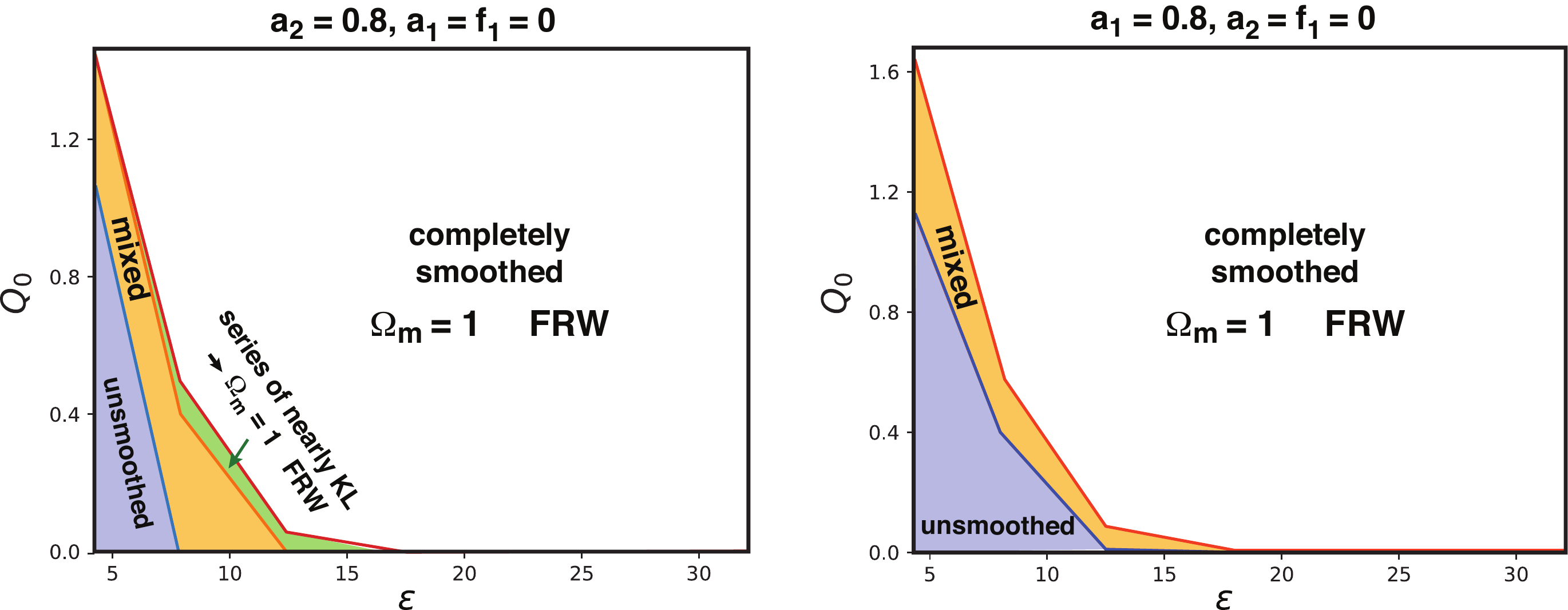}
\end{center}
\caption{Phase Diagram III (left) shows the final states reached for the cases in which the initial scalar field velocity is homogeneous $f_1 = 0$ but the off-diagonal components of $Z_{ab}^0$ are inhomogeneous ($a_2=0.8$ but $a_1 = 0$).
For Phase Diagram IV (right), only the on-diagonal components are highly inhomogeneous ($a_1 =  0.8$ but $a_2=0$).      
\label{fig:6}}
\end{figure*}

\vspace{0.1in}
\noindent
{\sc Phase Diagrams III and IV:} The third phase diagram (Fig.~\ref{fig:6}, left) corresponds to initial conditions in which the only initial inhomogeneity is due to  the spatially-varying off-diagonal components of $Z_{ab}^0$ in Eq.~(\ref{zic})  (the terms proportional to $a_2$). The initial scalar field velocity distribution is homogeneous ($f_1=0$) and the diagonal components of $Z_{ab}^0$ proportional to $a_1$ are also zero.  Here we find that, as in Phase Diagram I (Fig. 1),  the entire range of $Q_0 \ge 0$ and $\varepsilon \gtrsim 13$ converges directly to the $\Omega_m=1$ FRW dynamical attractor.  That is, unlike the previous example, introducing inhomogeneity by making $a_2$ non-zero does not reduce the robust smoothing and flattening behavior over the range of parameters relevant for cyclic and bouncing models. 
%
%
\begin{figure*}[t!]
\begin{center}
\includegraphics[width=6.00in,angle=-0]{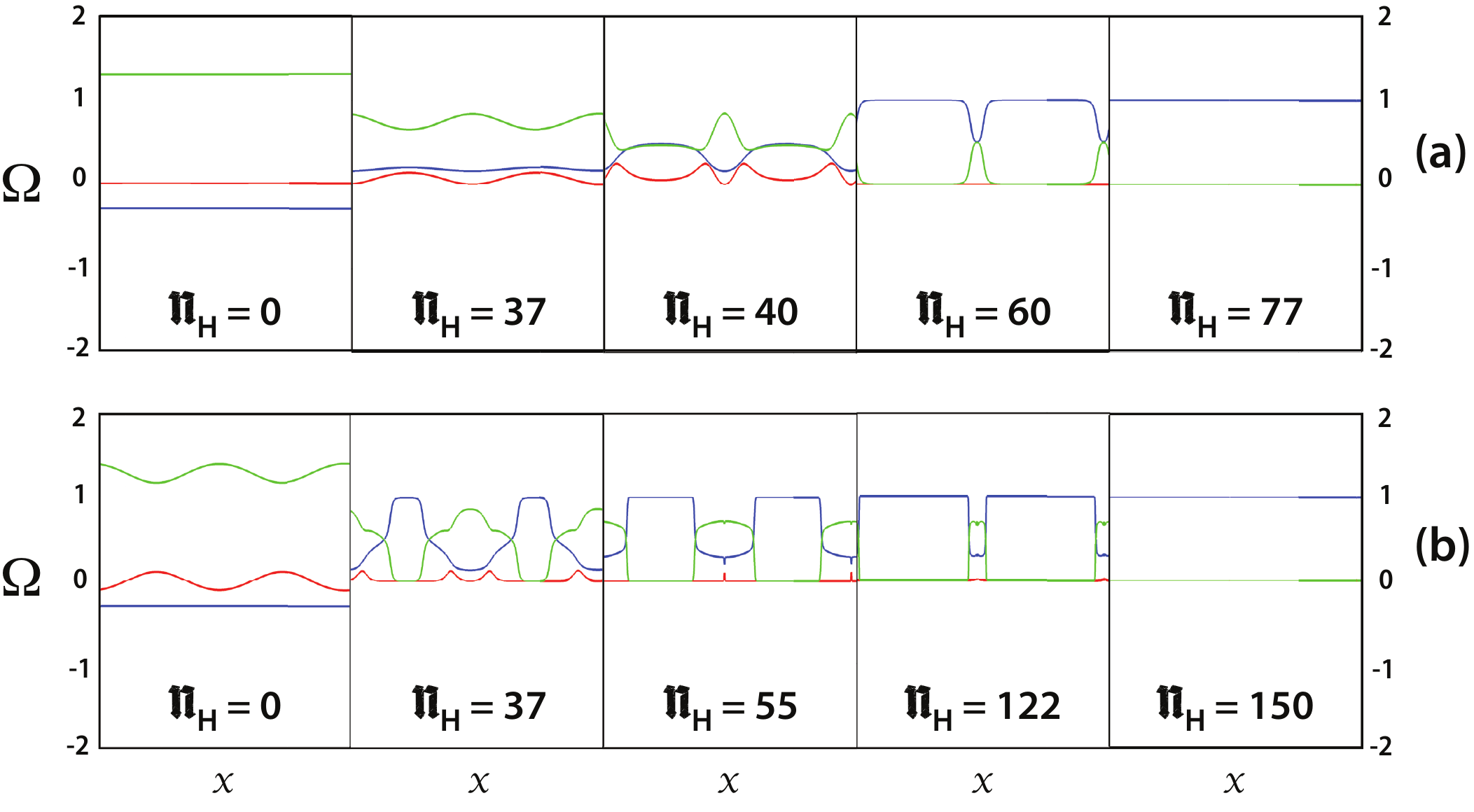}
\end{center}
\caption{Series of snapshots showing the evolution of the normalized energy density 
in matter $\Omega_m$
(blue), curvature $\Omega_k$ (red) and
shear $\Omega_s$ (green) 
for $0\le x \le 2 \pi$ for two cases with $\epsilon=13$ and $Q_0=0$ in which the inhomogeneity is solely
an off-diagonal inhomogeneous contribution to $Z_{ab}^0$: (a) $a_2=0.01$ and $f_1=a_1=0$; and (b) $a_2=1.0$ and $f_1=a_1=0$.  The first evolves through a series of nearly homogeneous  KL stages, then a series of inhomogeneous and mixed stages  before converging to a final homogenous  $\Omega_m=1$ FRW dynamical attractor; the second case begins significantly inhomogeneous (large $a_2$) and passes through a long sequence of different mixed states that include spikes (such as the one visible near the center of the lower panel labeled ${\mathfrak N}_H =55$) before reaching the final homogenous  $\Omega_m=1$ FRW dynamical attractor.  
\label{fig:7}}
\end{figure*} 

\begin{figure}[thb!]
\begin{center}
\includegraphics[width=4.750in]{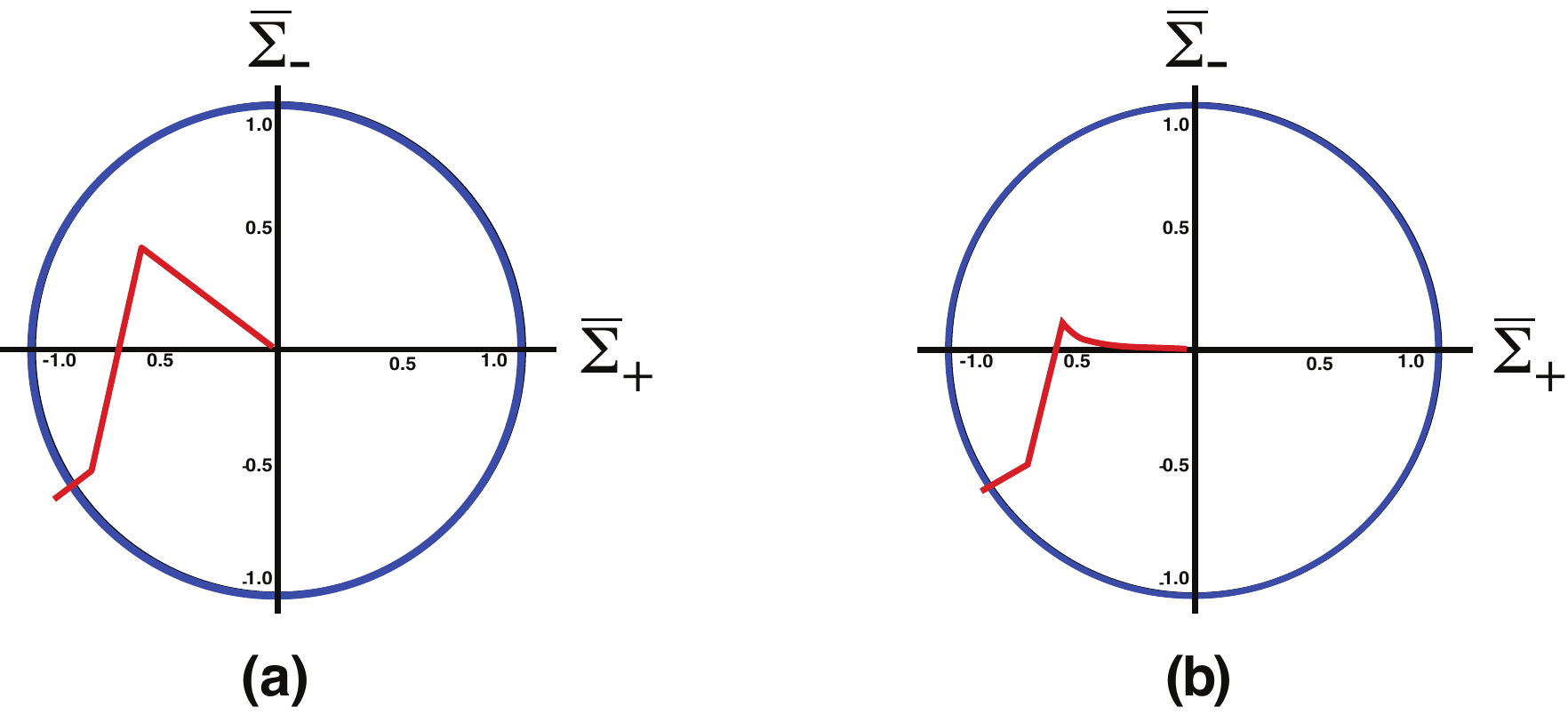}
\end{center}
\caption{The state space orbits comparing worldlines at
$x=\pi$ for the two models in Fig.~\ref{fig:7}: (a) one that passes through a series of nearly homogeneous KL stages but keeps veering away until finally converging to  the  $\Omega_m=1$ FRW dynamical attractor;  and (b) one that evolves through a different sequences of mixed states before converging to the attractor.  The point $x=\pi$ lies in a region that goes through multiple stages in both cases.  
 \label{fig:8}}
\end{figure}

In fact, a curious feature occurs for examples in the slivers of the phase diagram where  $\varepsilon \lesssim 13$ and $Q_0$ is near zero.   This is the range that leads to homogeneous KL behavior in Phase Diagram I.  When $a_2$ is non-zero and $\lesssim {\cal O}(0.1)$, we find instead that, beginning from a nearly homogeneous state,  the evolution first approaches a nearly homogeneous KL fixed point but then veers away and goes through a series of further nearly KL stages before converging on a $\Omega_m=1$ FRW dynamical attractor solution.  This is  illustrated in the upper panels shown in Fig.~\ref{fig:7} and  by the state space orbit diagram in Fig.~\ref{fig:8} (left).  
Effectively, this means that non-zero $a_2$ actually enlarges the phase space region that converges to the attractor solution.  As $a_2$ is increased by another order of magnitude, the  initial state is significantly inhomogeneous compared to the small $a_2$ case;  nevertheless, after evolving through a different sequence of stages, it also converges to the homogeneous $\Omega_m=1$ FRW dynamical attractor, as illustrated in the lower panels shown in Fig.~\ref{fig:7} space orbit diagram in Fig.~\ref{fig:8} (left).   (In the slivers of the phase diagram labeled as undergoing a series of nearly KL stages or as ending with mixed regions, the numerical evolution includes some spikes of negligible extent in $x$, as described in Ref.~\cite{Garfinkle:2008ei}.  An example is visible in the middle of the snapshot labeled $\mathfrak{N}_H=55$ in the lower row of Fig.~\ref{fig:7}; although not apparent in the subsequent snaphots, its presence can be identified in higher resolution runs.)
 
The fourth phase diagram (Fig.~\ref{fig:6}, right) is the complementary case where the non-zero inhomogeneous components of $Z_{ab}^0$ are on the diagonal ($a_1 \ne 0$) and the off-diagonal components are zero ($a_2=0$).  As in Phase Diagram II, the entire range of $Q_0 \ge 0$ and $\varepsilon \gtrsim 13$ converges directly to the $\Omega_m=1$ FRW dynamical attractor.  For  the corner of the phase diagram where $a_1 = {\cal O}(1)$, as in Fig.~\ref{fig:6} (right) and $\varepsilon \lesssim 13$,  the result is  a mixed state in which an exponentially large fraction of space time converges to the FRW dynamical attractor but there are also (cosmologically irrelevant) infinitesimal regions with spatially varying KL behavior.  For yet smaller $\varepsilon$, the spacetime is not smoothed.


\begin{figure*}[b!]
\begin{center}
\includegraphics[width=3.25in,angle=-0]{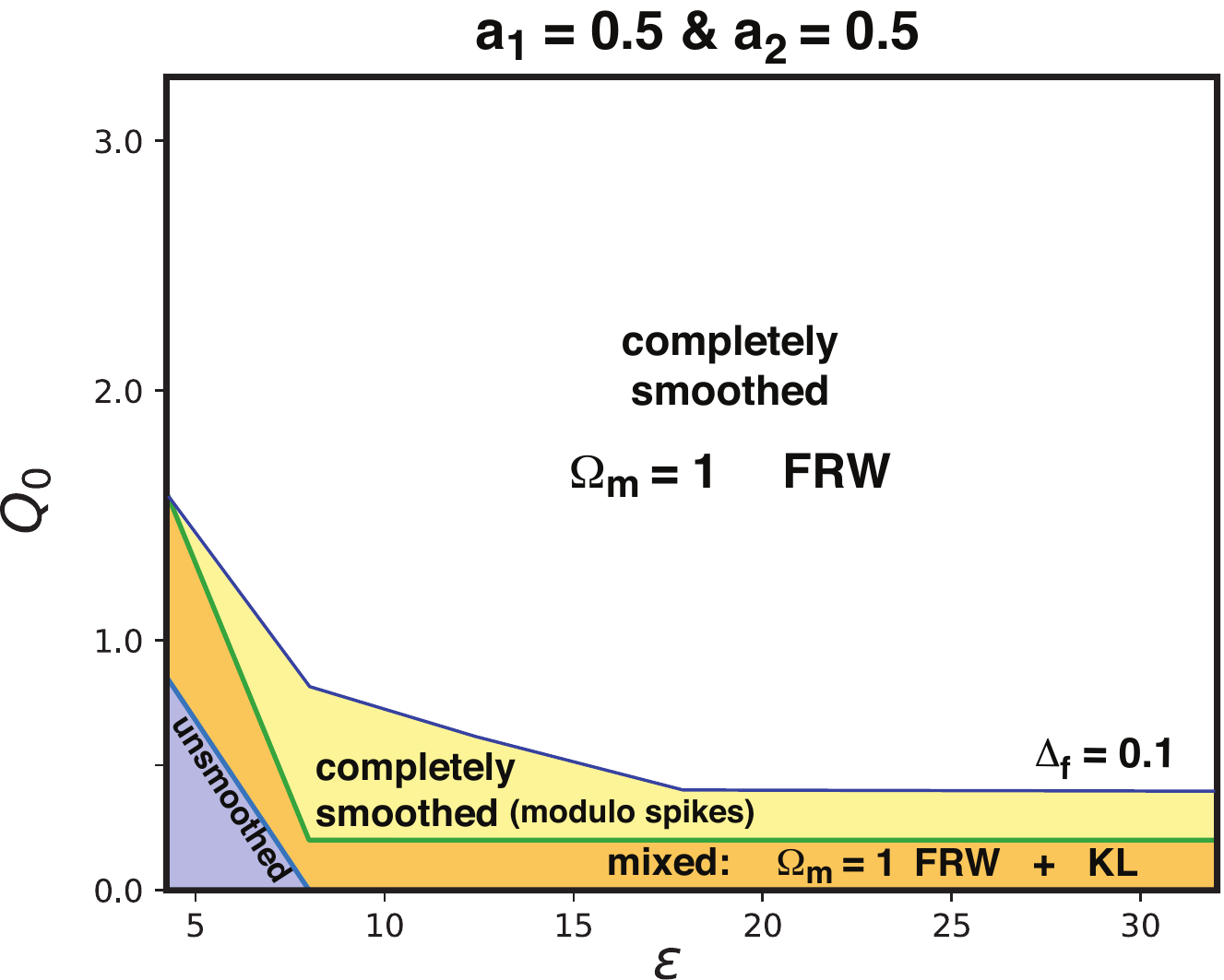}
\end{center}
\caption{Phase Diagram V shows the final states reached when a combination of inhomogeneities is considered (that is, is $f_1$, $a_1$ and $a_2$ are all non-zero).     The entire region relevant to cyclic and bouncing cosmology models --
$\varepsilon \gtrsim 13$ and $Q_0\ge 0$ converges completely or to an exponential degree (as measured by proper volume) to the desired smooth, anisotropic and flat dynamical attractor solution.
\label{fig:10}}
\end{figure*} 

\vspace{.1in}
\noindent
{\sc Phase Diagrams V:} Finally, we present a simplified phase diagram corresponding to cases in which  any combination of inhomogeneities (non-zero $f_1$, $a_1$ and $a_2$ up to ${\cal O}(1)$) is considered (Fig.~\ref{fig:10}). 
 This diagram encapsulates the take-away lesson regarding the remarkable robustness of slow contraction:  
{\it  The entire regime of practical relevance to cyclic and bouncing cosmology models} -- that is, having
 $\varepsilon \gtrsim 13$ (the condition for sufficiently rapid smoothing) and initial mean scalar field velocity  at rest or evolving down the potential $V(\phi)$ for all $x$)--
{\it converges completely or to an exponential degree (as measured by proper volume) to the desired smooth and flat FRW dynamical attractor solution with $\Omega_m=1$ and $\Omega_s=\Omega_k=0$.} 
(In the Appendix ~\ref{sec:app-convergence}, we describe our tests for numerical convergence and demonstrate that the tests are well-satisfied in those regions of the phase diagram that completely smooth.  The same is found to be true in the  regions of spacetime that directly smooth without spikes in the case of mixed final states.  More complex results are found in regions with spikey behavior.)
In this and in all the diagrams above, we allow highly non-perturbative initial conditions, but we restrict them  to be less than or ${\cal O}(1)$, which is at a level at or beyond what would be naturally expected as plausible initial conditions. Pushing parameters beyond the values considered here will move boundaries to slightly higher values of $\varepsilon$, say; but there  always remains a substantial range of the phase diagram that converges to the desired smooth and flat FRW dynamical attractor solution with $\Omega_m=1$ and $\Omega_s=\Omega_k=0$.   In particular, models with $\varepsilon >50$ (or $M< m_{\rm Pl}/50$ in Eq.~(\ref{potential}) where $m_{\rm Pl}$ is the reduced Planck mass) are plausible  in microphysical models,  and they are exponentially more powerful and rapid in  drawing much broader ranges of initial conditions to the dynamical attractor solution compared  to the already-robust cases considered here.

\begin{figure*}[tb!]
\begin{center}
\includegraphics[width=4.00in,angle=-0]{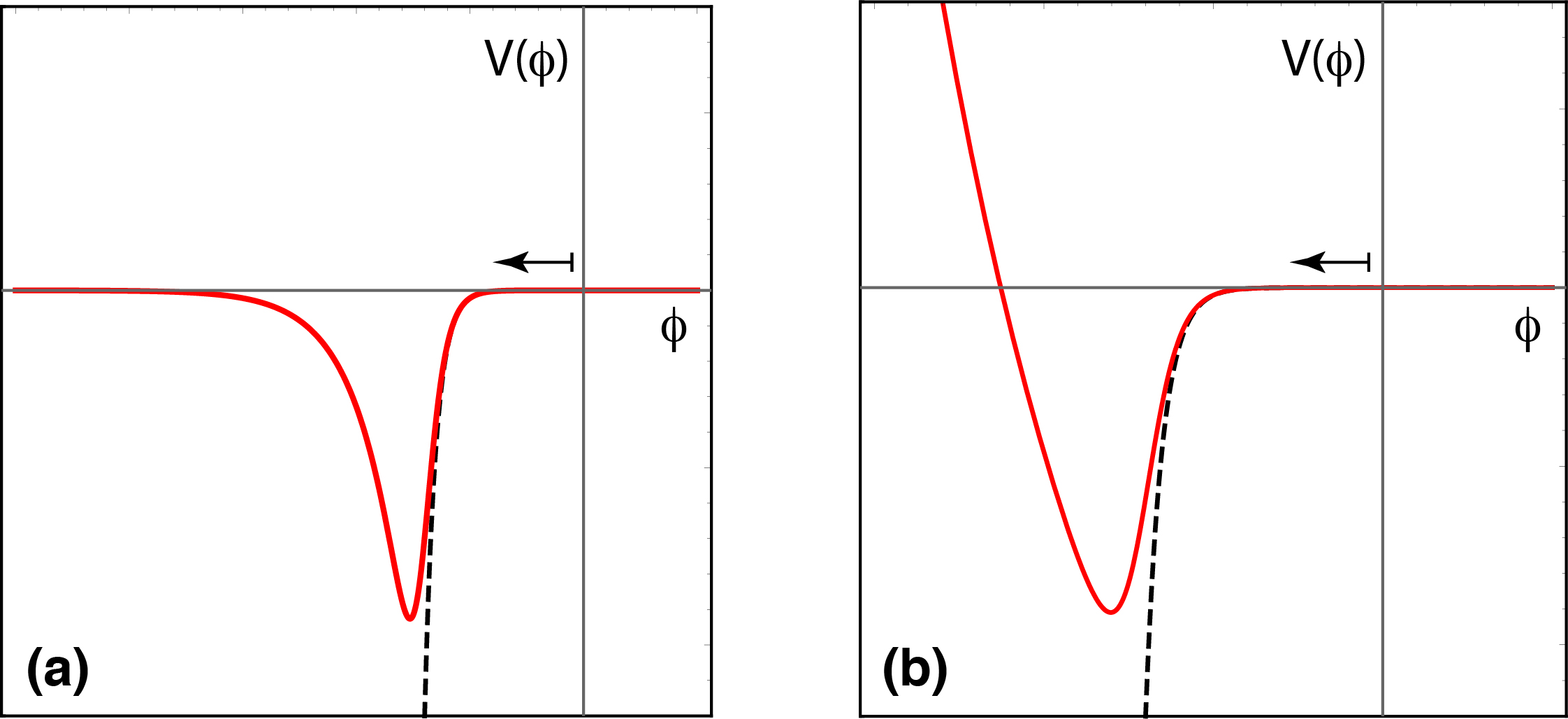}
\end{center}
\caption{Two types of potentials $V(\phi)$ (solid curves) used to test the robustness of slow contraction  in smoothing and flattening spacetime.  The scalar field evolves from right to left in each case (see arrow), first encountering a finite range of $V(\phi)$ that closely matches the negative exponential potential (dotted lines) considered in the phase diagrams above with $\varepsilon=50$.  From that point onward (after just a few $e$-folds of contraction),  the potentials deviate significantly: (a) the negative potential reaches a minimum and then approaches zero from below; and (b) the negative potential reaches a minimum and rises rapidly above zero with a shape that is super-exponential ({\it e.g.}, $\phi^4 \, {\rm exp} \,(\alpha \phi)$ for some constant $\alpha$). 
\label{fig:11}}
\end{figure*}

As a further test of robustness, a series of simulations were performed in which the scalar field potential $V(\phi)$ significantly deviates away from the simple negative potential  (Eq.~(\ref{pot}) that was assumed in the phase diagrams above.  The plots of $V(\phi)$ in Fig.~\ref{fig:11} illustrate two variations that were explored. 
%
%
\begin{figure*}[htb!]
\begin{center}
\includegraphics[width=5.75in,angle=-0]{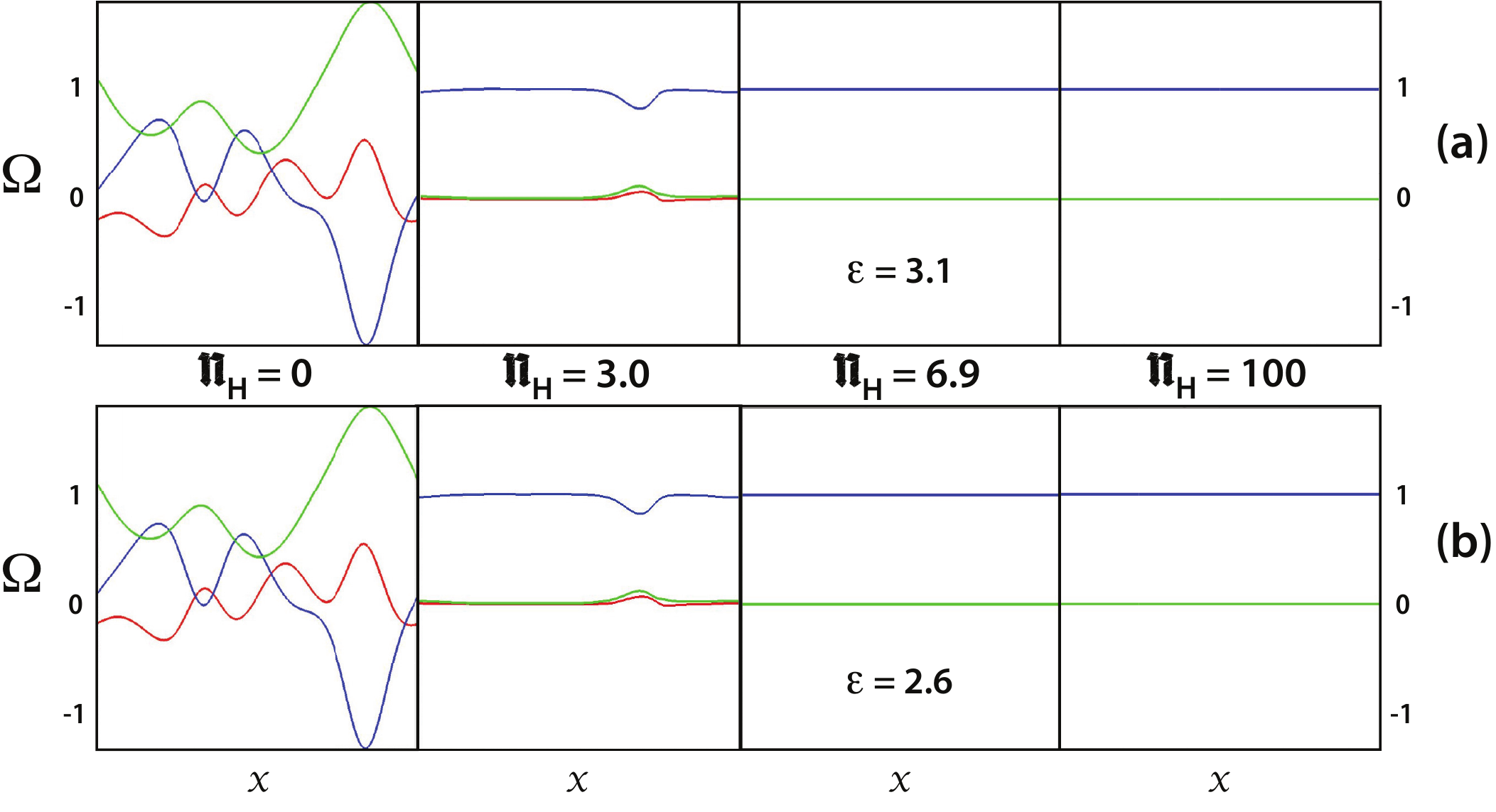}
\end{center}
\caption{Series of snapshots showing the evolution of the normalized energy density 
in matter $\Omega_m$
(blue), curvature $\Omega_k$ (red) and
shear $\Omega_s$ (green) 
for $0\le x \le 2 \pi$ for the two types of potentials described in Fig.~\ref{fig:11}.  In the top example, $\varepsilon_{\rm eff} \rightarrow 0$ as $\phi \rightarrow -\infty$; in the bottom example, $\varepsilon_{\rm eff}$ falls  below three as the $\phi$ climbs the steeply rising positive part of the potential.  In either case, the smoothing and flattening created by the initial slow contraction phase is maintained.
\label{fig:12}}
\end{figure*} 
%
%
\begin{figure}[htb!]
\begin{center}
\includegraphics[width=2.50in]{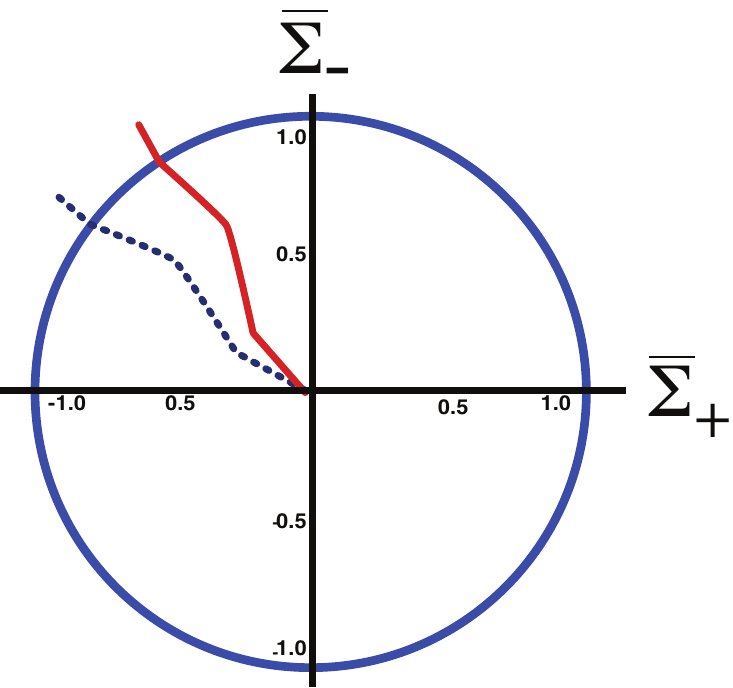}
\end{center}
\caption{The state space orbits comparing worldlines at
$x=3\pi/2$ for the two models in Fig.~\ref{fig:12}: (a) as $\phi \rightarrow -\infty$, $V(\phi)$ approaches zero from below (red, solid);
and (b) as $\phi \rightarrow -\infty$, $V(\phi)$ rises above zero and increases super-exponentially (blue, dotted). In both cases, the orbits converge rapidly to the center during the early slow contraction phase and remain there when the potential sharply deviates from negative exponential and the slow contraction phase ends.  
\label{fig:13}}
\end{figure}
In each case the scalar field evolves from right to left beginning with a segment that closely matches the negative exponential potential Eq.~(\ref{pot}).  During this first stage, there is a short period of slow contraction with effective equation-of-state $\varepsilon_{\rm eff} \rightarrow \varepsilon$, during which the large initial inhomogeneities are substantially smoothed and flattened as above. However, as $\phi \rightarrow -\infty$ (that is, evolving to the left), the potentials diverge and  $\varepsilon_{\rm eff}$ decreases rapidly and significantly.  

In the first case, the negative potential reaches a minimum and then rises to approach zero as $\phi \rightarrow -\infty$. The equation-of-state $\varepsilon_{\rm eff} \rightarrow 3$.  Despite this significant change, the early stage of slow contraction is rapid and powerful enough that the spacetime remains smooth and flat, as shown in the top panel of Fig.~\ref{fig:12} and by  the red solid trajectory in the state space orbit plot shown in Fig.~\ref{fig:13}.

\begin{figure}[tb!]
\begin{center}
\includegraphics[width=2.75in]{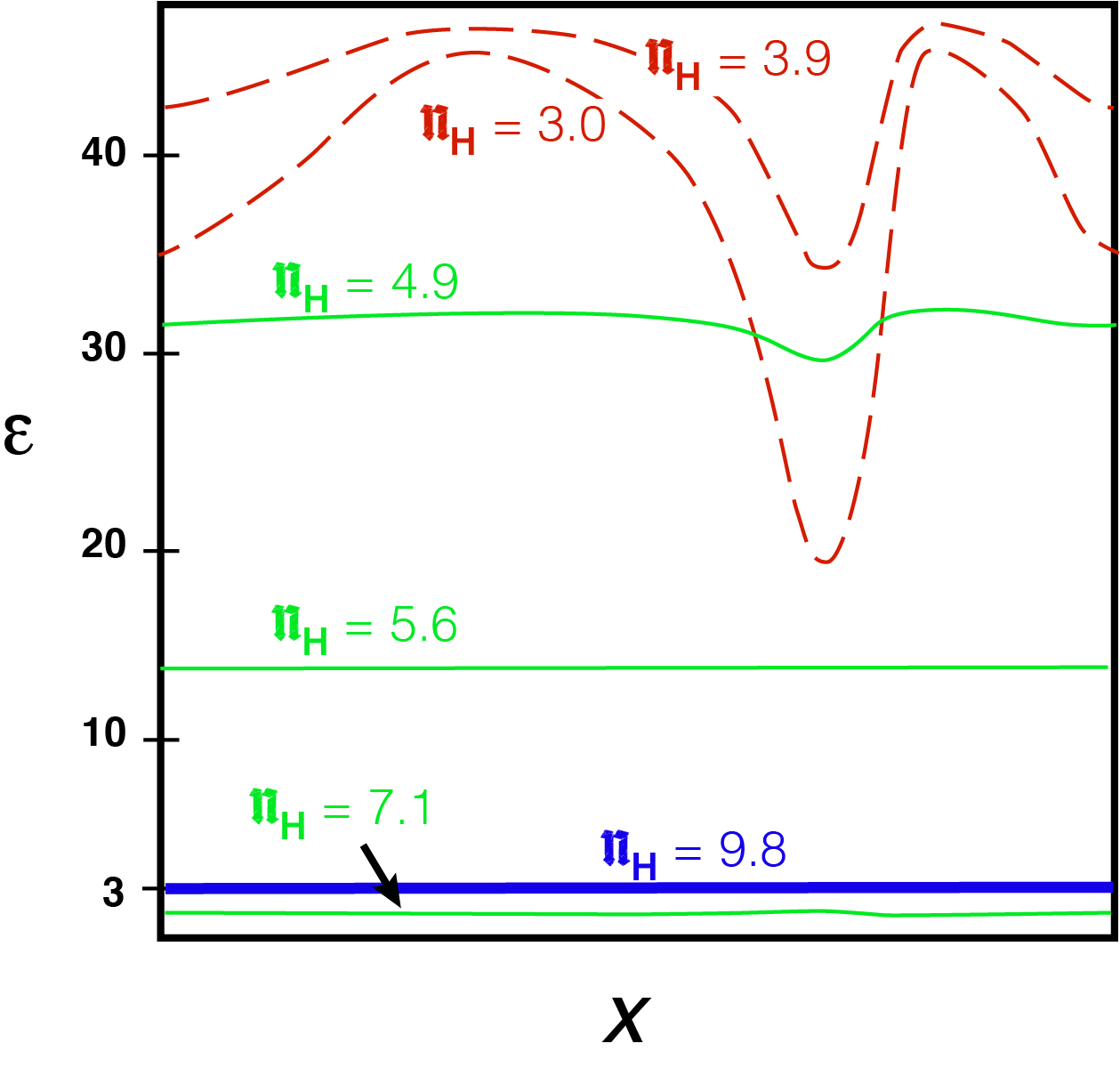}
\end{center}
\caption{A plot showing the effective equation-of-state $\varepsilon_{\rm eff}(x)$ at a sequence of times for the case shown in Fig.~\ref{fig:11}b in which $V(\phi)$ rises above zero and increases super-exponentially.  For early times when $V<0$ and exponentially decreasing   $\varepsilon_{\rm eff}$  rises to a large positive value (red dashed curves for which the number of e-folds of contraction are $n_H=3.0$ and~$3.9$).  At later times when $V>0$ and super-exponentially decreasing, the spacetime and $\varepsilon$ become nearly uniform and $\varepsilon_{\rm eff}$ decreases and falls below $\varepsilon=3$ (sequence of green thin solid curves with progressively increasing values of $n_H$).  After reaching a minimum at $n_H=7.1$, $\varepsilon_{\rm eff}$ increases and approaches the homogeneous FRW fixed point with  $\varepsilon =3$ (blue solid thick curve). 
\label{fig:14}}
\end{figure}

In the second  case, the negative potential reaches a minimum and then rises super-exponentially rapidly above zero; {\it e.g.,} $V(\phi) \propto \phi^4 \, {\rm exp} \,(\alpha \phi)$ as $\phi \rightarrow -\infty$.  The super-exponential behavior is artificially introduced  to force the equation-of-state  to reach $\varepsilon_{\rm eff}<3$, as shown in Fig.~\ref{fig:14}. Note that the potential  is not well-motivated  physically and does not correspond to any bouncing or cyclic picture; rather, it is specifically introduced as an extreme test of robustness.  Even in this case, the early stage of slow contraction is rapid and powerful enough that the spacetime remains smooth and flat,  as shown in the bottom panel of Fig.~\ref{fig:12} and by  the blue dotted trajectory in the state space orbit plot shown in  Fig.~\ref{fig:13}.

We have constructed examples of even more steeply rising, super-exponential potentials in which $\varepsilon_{\rm eff}$ not only falls below three, but also well below zero.  In this case, some small deviations from perfect homogeneity appear for some ranges of $x$ (a kind of mixed state) as the field climbs up the steep positive wall, though even then smoothness and flatness is retained for most of the range of $x$.

\section{Analytic approximation}
\label{sec:analytics}
 
 The key result of our numerical analysis is that slow contraction is an astonishingly robust smoothing mechanism: for most initial conditions, the scalar field energy density rapidly homogenizes and quickly becomes the dominant component ($\Omega_m\equiv1$), driving the geometry to a spatially-flat, homogeneous, and isotropic (FRW) spacetime everywhere and leading to {\it complete} smoothing for all $x$.   For extreme initial conditions that do not result in a completely smoothed FRW spacetime,  the end state is either a mixed that is smooth almost everywhere, as measured by co-moving volume, or  a homogeneous but anisotropic spacetime described by a `Kasner-like' (KL) metric. 
 To conclude our study, we complement our numerical computation with an analysis in which we demonstrate that the homogeneous end states found numerically correspond to the {\it only possible} stable critical points for a contracting universe.
 
 \subsection{Ultra-local limit} 
 
{\color{black} Notably,  in all the cases we studied, the evolution towards the homogeneous end states found numerically approach the so-called {\it ultra-local} limit, in which all terms that involve gradients become negligible in the evolution and constraint equations, {\it i.e.},
 \begin{equation}
\bar{E}_a{}^i \to 0, \bar{A}_a \to 0, \bar{S}_a \to 0,
\end{equation}
as the universe contracts ($t\to-\infty$).  
Most remarkably, ultra-locality is generically reached rapidly even in those cases where the initial conditions include large gradients, as is apparent from the plot of $\Omega_k$ in the $\mathfrak{N}_H=0$ panels shown in Figs.~\ref{fig:4},  \ref{fig:7} (lower panel) and \ref{fig:12}.  Note that, by choosing the spatial metric on the initial $t_0$-hypersurface to be conformally flat, $\Omega_k$ is a direct measure of spacetime gradients. }


 In the ultra-local limit, the evolution and constraint equations~(\ref{eq-E-ai-Hn}, \ref{eq-A-a}, \ref{eq-S-a-Hn} and \ref{constraintJ}-\ref{constraintCOM}) for $\bar{E}_a{}^i, \bar{A}_a$ and $\bar{S}_a$, respectively, are automatically satisfied, while the remaining system of non-trivial evolution and constraint equations dramatically simplifies, leaving a system of first-order ordinary differential equations (ODEs) for twelve metric and two matter variables constrained by three algebraic relations, as opposed to the original set of coupled partial differential equations with twenty-four metric and five matter variables.

Most importantly, in the ultra-local limit the momentum constraint~\eqref{constraintC},
\begin{equation}
\label{constraintC-ul}
\epsilon_a{}^{bc} \bar{n}_b{}^d \bar{\Sigma}_{cd} = 0,
\end{equation}
is equivalent to demanding that the shear tensor $\bar{\Sigma}_{ab}$ and (the symmetric part of) the intrinsic curvature tensor $\bar{n}_{ab}$ commute, which means that $\bar{\Sigma}_{ab}$ and $\bar{n}_{ab}$ share the same {\it eigenvectors}. Furthermore, 
as shown in the Appendix~\ref{sec:app-eigensystem}, the eigenvectors at time $t_0$ remain eigenvectors at later times.  The dynamics in the ultra-local limit, therefore, is completely determined by the  evolution of the {\it eigenvalues} of $\bar{\Sigma}_{ab}$ and $\bar{n}_{ab}$.

As a consequence, the twelve coupled ODEs for the metric variables $\bar{\Sigma}_{ab}$ and $\bar{n}_{ab}$,
\begin{eqnarray}
\label{eq-sigma-ab-ul}
\dot{\bar{\Sigma}}_{ab} &=& -\Big( 3{\cal N} - 1 \Big) \bar{\Sigma}_{ab} - {\cal N} \Big( 
2 \bar{n}^c{}_{\langle a} \bar{n}_{b\rangle c } - \bar{n}^c{}_c \bar{n}_{\langle ab \rangle} \Big),\\
\label{eq-n-ab-ul}
\dot{\bar{n}}_{ab} &=&  -\Big( {\cal N} - 1 \Big) \bar{n}_{ab} + 2\, {\cal N} \,  \bar{n}^c{}_{(a} \bar{\Sigma}_{b )c} ,
\end{eqnarray}
 can be replaced by six ODEs for the corresponding eigenvalues $\sigma_i, \nu_i$ ($i=1,2,3$), such that
 the dynamical system is fully described by the following set of six evolution equations
\begin{eqnarray}
\label{eq-sigma1-ul}
\dot{\sigma}_1 &=&   \Big(1 - 3{\cal N} \Big) \sigma _1
- {\textstyle \frac13 }  \, {\cal N}\left(   \big(2\nu_1 - \nu_2 - \nu_3 \big) \nu_1 - \big(  \nu_2- \nu_3  \big)^2 \right)
,\\
\dot{\sigma}_2 &=&   \Big(1 - 3{\cal N} \Big) \sigma _2
- {\textstyle \frac13 }  \, {\cal N}\left(   \big(2\nu_2 - \nu_1 - \nu_3 \big) \nu_2 -  \big(  \nu_1- \nu_3  \big)^2
 \right)
 ,\\
\label{nu1-ev-ul-app}
\dot{\nu}_1 &=& \Big( 1 + {\cal N} \big(   2   \sigma_1 - 1 \big) \Big) \nu_1  
,\\
\dot{\nu}_2 &=&  \Big( 1 + {\cal N} \big(   2   \sigma_2 - 1 \big) \Big) \nu_2  
,\\
\dot{\nu}_3 &=& \Big( 1 - {\cal N} \big( 2\sigma_1 + 2\sigma_2  + 1 \big)   \Big) \nu_3   
,\\
\label{eq-w-ul}
\dot{\bar{W}}&=& -\Big( 3{\cal N} - 1 \Big) \bar{W}  +  \sqrt{2\varepsilon} \,{\cal N} \,\bar{V} 
 ,
\end{eqnarray}
subject to the Hamiltonian constraint
\begin{equation}
\label{constraintG-ul}
{\cal N}^{-1}  - {\textstyle \frac16 } \Big(\nu_1^2+\nu_2^2+\nu_3^2\Big)
+ {\textstyle \frac{1}{12} } \Big( \nu_1+\nu_2+\nu_3 \Big)^2 
-  \Big( \sigma_1^2 + \sigma_1\sigma_2 + \sigma_2^2 \Big)
- {\textstyle \frac12 } \bar{W}^2  = 0,
\end{equation}
and the ultra-local limit of the Hubble-normalized lapse equation~\eqref{Neqn},
\begin{equation}
\label{Neqn-limit}
{\cal N}^{-1} = 1+  {\textstyle \frac23 }\Big( \sigma_1^2 + \sigma_1\sigma_2 + \sigma_2^2  \Big) + {\textstyle \frac13} \Big( \bar{W}^2   - \bar{V}(\phi) \Big).
\end{equation}
As before, dot denotes differentiation with respect to (coordinate) time $t$, as defined in Eq.~\eqref{timechoice}.
The third shear eigenvalue $\sigma_3$ could be eliminated since the trace-freeness of the shear tensor implies that the three eigenvalues must sum to zero.
In addition, we substituted in Eq.~(\ref{eq-w-ul})
\begin{equation}
\frac{\bar{V}_{,\phi}}{\bar{V}} = - \sqrt{2\varepsilon},
\end{equation}
 to eliminate the term proportional to  $\bar{V}_{,\phi}$ in Eq.~(\ref{eq-w-Hn}) (and, hence, eliminate any explicit $\phi$-dependence from the equations above).

%

\subsection{Critical point solutions}

As detailed above, our goal is to show that the homogenous end states we identified numerically are the stable attractor solutions of the underlying system of evolution and constraint equations above. Since any attractor is a (stationary) critical point, we begin with identifying all critical points of the system as given by Eqs.~(\ref{eq-sigma1-ul}-\ref{eq-w-ul}).
 We list the complete set of (seven) critical point solutions of Eqs.~(\ref{eq-sigma1-ul}-\ref{eq-w-ul}) in Table~\ref{table-1}.
%
\begin{table}[tbp]
\centering
\renewcommand{\arraystretch}{2}
\begin{tabular}{ |c|c| c|c|c|c|c| c|  }
 \hline
 ${\cal N}$ & $\sigma_1$ & $\sigma_2$ & $\nu_1$ & $\nu_2$ & $\nu_3$ & ${\textstyle \frac12 }\bar{W}^2$ & $\bar{V}$   
 \\ \hline  \hline
  ${\textstyle \frac{1}{\varepsilon}}$ & 0 & 0 & 0 & 0 & 0 &$\varepsilon>0$  & $3-\varepsilon$   
  \\ \hline
 ${\textstyle \frac13 }$ & $\neq0$ & $\neq0$ & 0 & 0 & 0 & {\small $3-(\sigma_1^2 + \sigma_1\sigma_2 + \sigma_2^2)$} & 0 
 \\ \hline
 ${\textstyle \frac13 }$ & $-1$ & $-1$ & $\neq0$ & $\nu_1$ & 0 & 0 & 0 
 \\ \hline
 \hline
\rule{0pt}{17.5pt}
 1 & 0 &0 & $\pm{\textstyle 2\sqrt{ \frac{1}{\varepsilon}-1}}$ & $\nu_1$  &  $\nu_1$ & ${\textstyle \frac{1}{\varepsilon}}>1$& $\bar{W}^2$   \\[1ex] \hline
 1 & 0 &0 & $\neq0$ & 0  & $\nu_1$ & ${\textstyle \frac12 }{\textstyle \varepsilon} = {\textstyle \frac12 }$& $\bar{W}^2$   \\ \hline
  ${\textstyle \frac{\varepsilon+8}{9\varepsilon}}$ & ${\textstyle \frac{4-4\varepsilon}{\varepsilon+8}}$ & ${\textstyle \frac{2\varepsilon-2}{\varepsilon+8}}$ & $\pm  {\textstyle 6\frac{\sqrt{(1-\varepsilon)(\varepsilon-4)}}{\varepsilon+8}}$ & 0 & 0 & $ 1<{\textstyle \frac{81\varepsilon}{(\varepsilon+8)^2}}<4$ & $  {\textstyle \frac{54 \cdot (4-\varepsilon)}{(\varepsilon+8)^2}}$   \\[1ex] \hline
 ${\textstyle \frac{\varepsilon+2}{3\varepsilon}<1}$& ${\textstyle \frac{1-\varepsilon}{2+\varepsilon}}$ & 
  ${\textstyle \frac{2\varepsilon-2}{2+\varepsilon}}$ & ${\textstyle \pm3 \frac{\sqrt{\varepsilon-1}}{\varepsilon+2}}$ & 0 & $-\nu_1$ & ${\textstyle \frac{9\varepsilon}{(2+\varepsilon)^2}}$ & ${\textstyle \frac{18\varepsilon}{(2+\varepsilon)^2}}$ \\[1ex] 
 \hline
\end{tabular}
\caption{Critical point solutions corresponding to the autonomous system of ordinary differential equations~(\ref{eq-sigma1-ul}-\ref{eq-w-ul}) describing the evolution in the ultra-local limit.
\label{table-1}}
\end{table}
%

If the potential energy density $V(\phi)$ is negative, as needed to drive slow contraction (see Sec.~\ref{sec_intro} and Refs.~\cite{Khoury:2001wf,Steinhardt:2001st,Ijjas:2018qbo,Ijjas:2019pyf}), there exists only the three distinct critical point solutions listed in the first three lines of Table~\ref{table-1}:
\begin{itemize}
\item[-] flat, homogeneous, and isotropic (FRW) scaling solution: 
\item[-] flat, homogeneous, and anisotropic  (Kasner-like with $\Omega_m \ge 0$) solution; and
\item[-] curved, homogeneous, and anisotropic ($\Omega_m = 0$) solution.
\end{itemize}
{\color{black} Since the remaining four critical points} listed in rows 4-7 of Table~\ref{table-1} only exists for positive potentials, they cannot trigger a phase of slow contraction and lie outside of the scope of this paper. For this reason, we do not further consider them here.


\subsection{Stability of critical point solutions}

A critical point is an {\it attractor} solution if it is stable to small perturbations. Otherwise, it is a {\it repeller.} Accordingly, to decide which of the three critical points are attractor solutions, we linearize the system around each critical point and solve the resulting constant-coefficient system for the complete set of perturbation variables.  (Here, we only show the three sets of equations corresponding to the critical points. For the perturbed evolution and constraint equations around an arbitrary background, see  Appendix~\ref{sec:appPerturbedEqs}.)

\subsubsection{FRW (scaling) attractor solution with $\varepsilon >3$}

First, we linearize the system around the homogeneous and isotropic FRW  critical point
\begin{equation}
  \sigma_i = \nu_i= 0 \quad({\rm for\, all}\, i); 
  \quad {\cal N}^{-1} =  {\textstyle \frac12 }\bar{W}^2 = \varepsilon;\quad 
  {\bar V} = 3 - \varepsilon.
\end{equation}

The perturbed shear, spatial curvature and scalar-field matter decouple and obey the following evolution equations:
\begin{eqnarray}
\delta \dot{\sigma}_i &=&   \left(1 - \frac{3}{\varepsilon} \right) \delta \sigma _i,   \quad i=1,2;
\\
\label{nu1-ev-ul-app-p0}
\delta\dot{\nu}_j &=& \left( 1 - \frac{1}{\varepsilon} \right) \delta \nu_j,   \quad j=1,2,3;
\\
\label{eq-w-ul-p0}
\delta\dot{\bar{W}}&=& \left( 1-\frac{3}{\varepsilon} \right) \delta \bar{W}.
\end{eqnarray}
Solutions to the system admit a simple exponential scaling behavior,
\begin{eqnarray}
\delta \sigma_i, \delta \bar{W} \propto e^{ \big(1 - \frac{3}{\varepsilon} \big) t}, \quad \delta\nu_j \propto e^{ \big(1 - \frac{1}{\varepsilon} \big) t}.
\end{eqnarray}
It is immediately apparent that, for a contracting spacetime ($t\to-\infty$), the FRW scaling solution is a stable attractor for $\varepsilon>3$ and a repeller otherwise.

Notably, the FRW critical point solution recovers the well-known scaling attractor solution,
\begin{equation}
\label{FRW-scaling-sol}
a = (\tau/\tau_0)^{1/\varepsilon}, \phi = \sqrt{2/\varepsilon}\ln (\tau/\tau_0),
\end{equation}
(where $\tau$ is the proper FRW time), as often cited in the cosmology literature. In particular, the {\it eigenvalues} correspond to the so-called `Friedmann variables' commonly used to identify the scaling solution while assuming an FRW background. (For details, see the Appendix~\ref{sec:appFrVar}.)

\subsubsection{Kasner-like attractors and repellers}

Linearizing Eqs.~(\ref{eq-sigma1-ul}-\ref{eq-w-ul}) for the homogeneous and anisotropic Kasner-like critical point solution,
\begin{equation}
{\cal N} =  {\textstyle \frac13 },\quad  \sigma_1, \sigma_2\neq0, \quad \nu_1,\nu_2,\nu_3 =0, \quad 
{\textstyle \frac12 } \bar{W}^2  = 3-(\sigma_1^2 + \sigma_1\sigma_2 + \sigma_2^2), \quad {\bar V} = 0,
\end{equation}
the system reduces to two simple decoupled constant-coefficient matrix equations: one for the three spatial curvature {\it eigenvalue} variables,
\\
\begin{equation}
\renewcommand{\arraystretch}{1.3}
\left( \begin{array}{c} 
 \delta\dot{\nu}_1 \\ \delta\dot{\nu}_2 \\ \delta\dot{\nu}_3 
\end{array} \right)
= 
\frac23 \begin{pmatrix}
\sigma_1 + 1 & 0 & 0 \\
0 &   \sigma_2 + 1 & 0 \\
0 & 0 & 1 -  \sigma_1  -  \sigma_2
\end{pmatrix}
\left( \begin{array}{c} 
 \delta\nu_1 \\ \delta\nu_2 \\ \delta\nu_3 
\end{array} \right)
,\end{equation}
\\
and one for the shear {\it eigenvalue} and scalar-field matter variables,
\\
\begin{equation}
\renewcommand{\arraystretch}{1.2}
\left( \begin{array}{c} 
\delta \dot{\sigma}_1 \\ \delta \dot{\sigma}_2 \\ \delta\dot{\bar{W}}
\end{array} \right)
=
J^{-1}\begin{pmatrix}
0 & 0 & 0  \\
0 & 0 & 0  \\
0 & 0 & p
\end{pmatrix}
J
\left( \begin{array}{c} 
\delta \sigma_1 \\ \delta \sigma_2 \\ \delta \bar{W}
\end{array} \right), 
\end{equation}
where 
\begin{equation}
p\equiv \frac23 \left(3   - \sqrt{\varepsilon} \cdot \sqrt{3-(\sigma_1^2 + \sigma_1\sigma_2 + \sigma_2^2)} \right),
\end{equation}
and the matrix $J$ is given by
\begin{equation}
J=
\renewcommand{\arraystretch}{1.5}
\begin{pmatrix}
- \frac{\bar{W}}{2\sigma_1 +\sigma_2} &  -\frac{\sigma_1 + 2\sigma_2}{2\sigma_1 +\sigma_2 } & \frac{\sigma_1}{\bar{W}-\sqrt{2\varepsilon}}   \\
0 & 1 &  \frac{\sigma_2}{\bar{W}-\sqrt{2\varepsilon}}   \\
1 & 0 & 1
\end{pmatrix}.
\end{equation}
 
\begin{figure*}[htb!]
\begin{center}
\includegraphics[width=5.25in,angle=-0]{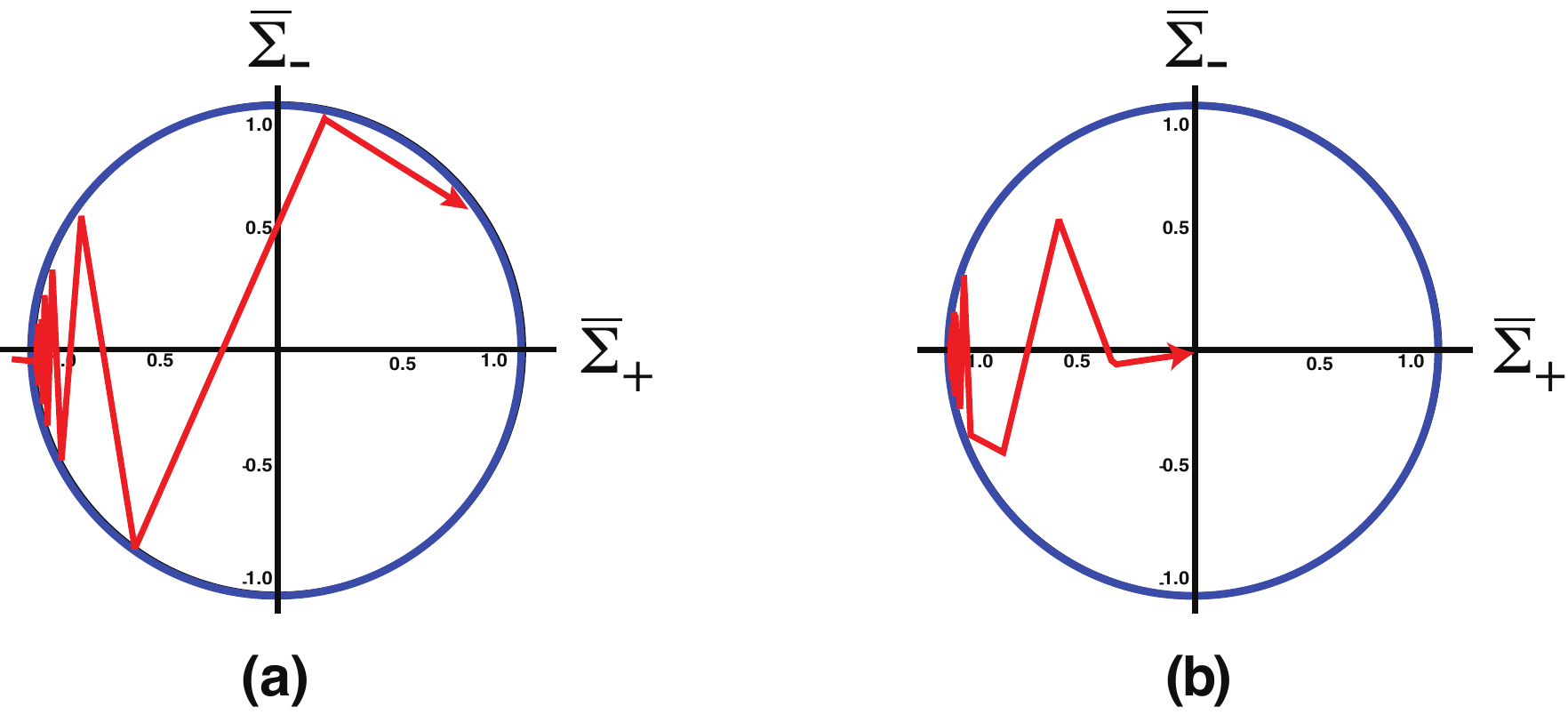}
\end{center}
\caption{State orbit plots for two cases: (a) starting with initial conditions $\bar{W}=0$, $\{\sigma_1, \sigma_2, \sigma_3 \} = \{ -1.5, -1.6, 3.1\} $ and $\{ \nu_1, \nu_2, \nu_3\} = \{0.2, 0.3 ,0\}$ and  $\varepsilon=0$ in Eq.~\eqref{eq-w-ul}, spacetime undergoes a series of bounces before converging to a homogeneous Kasner attractor solution with $\bar{W}=0$, $\{\sigma_1, \sigma_2, \sigma_3\} = \{ 1.8, -2.5, -1.6\} $ and $\{ \nu_1, \nu_2, \nu_3\} = \{0, 0,0\}$. 
and (b) starting with initial conditions $\bar{W}=0$, $\{\sigma_1, \sigma_2, \sigma_3 \} = \{ -1.02, -0.98, 2\} $ and $\{ \nu_1, \nu_2, \nu_3\} = \{0.2, 0.3 ,0\}$ and  $\varepsilon=6$, spacetime undergoes a different sequence of bounces before converging to a homogeneous and isotropic (spatially-flat) FRW attractor solution with $\bar{W}=\sqrt{6}$,  $\{\sigma_1, \sigma_2, \sigma_3\} = \{ 0, 0, 0\} $ and $\{ \nu_1, \nu_2, \nu_3\} = \{0, 0 ,0\}$.    
\label{fig:15}}
\end{figure*}

As before, solutions to the system admit a simple exponential scaling behavior:
\begin{align}
&\delta \nu_{1,2} \propto e^{\frac23 \big(1+\sigma_{1,2}\big) t}, \quad \delta\nu_3 \propto e^{\frac23 \big(1-\sigma_1-\sigma_2\big) t},  \\
&\delta \sigma_{1,2}, \delta \bar{W} \propto e^{\frac23 \big(3   - \sqrt{\varepsilon} \cdot \sqrt{3-(\sigma_1^2 + \sigma_1\sigma_2 + \sigma_2^2)} \big) t}.
\end{align}
Intriguingly, though, the Kasner-like solution {\it can} behave both as an attractor or a repeller in a contracting universe ($t\to-\infty$): if $\varepsilon =0$ and $\sigma_i>-1$ for $i=1,2,3$; or if $\varepsilon >0$, $\sigma_i>-1$ and  $(\sigma_1^2 + \sigma_2^2+\sigma_3^2)> 3- 9/\varepsilon $, the Kasner-like solution is an attractor. Otherwise, it is a repeller.
Examples for both behaviors are shown in Figure~\ref{fig:15}. 
%

In Case (a) on the left, the system converges to a Kasner-like state, while in Case (b) on the right,  the system is driven to the FRW scaling attractor solution. This second case is especially important since it makes apparent the difference to the well-known vacuum case (that is, pure gravity with no scalar field). In the pure vacuum case, the Kasner solution is the only stable attractor. In the presence of a scalar-field, though, reaching the Kasner-like attractor is only possible under very special initial conditions, namely, when the initial scalar field velocity is uphill (that is, in the direction that $V(\phi)$ increases). In this case, the scalar field's relative contribution to the total energy density ($\Omega_m$) becomes negligible and there is the same dynamical behavior as  in the vacuum case. But for cosmologically motivated initial conditions, as discussed in Ref.~\cite{Cook:2020oaj} and the previous section, the initial scalar field velocity is at rest or downhill; then one finds that  the scalar field energy density increases relative to the other components and  the FRW scaling solution becomes the only attractor.

An illustrative example for how large the set of initial data is that belongs to the basin of attraction of the FRW solution was given above in Figure~\ref{fig:7}. There the initial scalar field matter energy density was chosen so small that the system first approached a Kasner-like critical point, just as it would in the absence of matter. Then the system was dynamically driven away by the growing homogeneous spatial curvature to another Kasner-like critical points through a series of (mixmaster) bounces. Yet, as the scalar field's energy density continued to grow relative to other contributions,  all Kasner-like critical points became  repellers and the system eventually settled in the FRW attractor solution.

Finally, we note that this analysis also complements and generalizes our numerical study in that it enables us to follow the evolution under homogeneous but spatially curved ($\nu_j\neq0$ for at least one $j\in\{1,2,3\}$)  initial data. This is the one type of initial data that is excluded in our numerical study by assuming a conformally-flat spatial metric on the initial spacelike $t_0$-hypersurface.

\subsubsection{Curved, homogeneous and anisotropic repeller solutions}

Thirdly and lastly, we linearize the system~(\ref{eq-sigma1-ul}-\ref{eq-w-ul}) around the {\color{black} curved homogeneous and anisotropic} critical point solution,
\begin{eqnarray}
{\cal N} = {\textstyle \frac13 },\quad
\sigma_1 =  \sigma_2 = -1,\quad
\nu_1  =  \nu_2 \equiv \nu\neq0,\quad
\nu_3 =  0,\quad
 \bar{W}=0,\quad
 {\bar V}  = 0,
\end{eqnarray}
leading to a simple set of evolution equations:
\begin{equation}
\renewcommand{\arraystretch}{1.3}
\left( \begin{array}{c} 
\delta \dot{\sigma}_1 \\ \delta \dot{\sigma}_2 \\ \delta\dot{\nu}_1 \\ \delta\dot{\nu}_2 \\ \delta\dot{\nu}_3 \\ \delta\dot{\bar{W}}
\end{array} \right)
= 
\begin{pmatrix}
1 & 1 & -  {\textstyle \frac13 } \nu &  {\textstyle \frac13 } \nu & 0 & 0 \\
1 & 1 &  {\textstyle \frac13 } \nu &  -{\textstyle \frac13 } \nu & 0 & 0 \\
-{\textstyle \frac13 } \nu &  - \nu & 0 & 0 & -{\textstyle \frac19 } \nu^2 &0\\
- \nu & -{\textstyle \frac13 } \nu &   0 & 0 & -{\textstyle \frac19 } \nu^2 &0\\
0 & 0 & 0 & 0 & 2 & 0 \\
\sqrt{2\varepsilon} & \sqrt{2\varepsilon} & 0 & 0& \sqrt{2\varepsilon} \, {\textstyle \frac{\nu}{3} } & 0
\end{pmatrix}
\cdot
\left( \begin{array}{c} 
\delta \sigma_1 \\ \delta \sigma_2 \\ \delta\nu_1 \\ \delta\nu_2 \\ \delta\nu_3 \\ \delta \bar{W}
\end{array} \right).
\end{equation}
\\
%
%
For some initial perturbations $\delta \sigma_i^{0}, \delta \nu_j^{0}, \delta \bar{W}^{0}$,  the system admits the following solutions:
\begin{eqnarray}
\delta \sigma_1 &=& \frac12 \left( \delta \sigma_1^0 - \delta \sigma_2^0+ \Big(\delta \sigma_1^0 + \delta \sigma_2^0 \Big) e^{2 t } \right)
,\\
\delta \sigma_2 &=& \frac12 \left( \delta \sigma_2^0 - \delta \sigma_1^0+ \Big(\delta \sigma_1^0 + \delta \sigma_2^0 \Big) e^{2 t } \right)
,\\
\delta \nu_1 &=& \delta \nu_1^{0} 
+ \frac{ \nu}{3}  \Big( \delta \nu_1^{0} - \delta \nu_2^{0} \Big) t 
+  \frac{ \nu}{3}  \Big(  \delta \nu_1^{0} + \delta \nu_2^{0} 
+  {\textstyle \frac{\nu}{6} } \delta \nu_3^{0}\Big) \left(1 - e^{2 t } \right)
,\\
\delta \nu_2 &=& \delta \nu_2^{0} 
- \frac{ \nu}{3} \Big( \delta \nu_1^{0} - \delta \nu_2^{0} \Big)  t
+  \frac{ \nu}{3} \Big(  \delta \nu_1^{0} + \delta \nu_2^{0} 
+  {\textstyle \frac{\nu}{6} } \delta \nu_3^{0}\Big) \left(1 - e^{2 t } \right)
,\\
\delta \nu_3 &=& \delta \nu_3^{0}\, e^{2 t }
,\\
\delta \bar{W} &=& \delta \bar{W}^0 - \sqrt{\frac{\varepsilon}{2}} \Big( \delta \nu_1^{0}  + \delta \nu_2^{0}  + {\textstyle \frac{\nu}{3}} \delta \nu_3^{0}   \Big) \left( 1- e^{2 t } \right) 
.
\end{eqnarray}

Unlike before, as contraction proceeds, the dominant scaling behavior is {\it not} the exponential but the constant or power-law terms in the solutions. More exactly, as $t\to-\infty$, the shear and scalar field fluctuations converge to constants
\begin{equation}
\delta \sigma_1 \to {\textstyle \frac12} \left( \delta \sigma_1^0 - \delta \sigma_2^0\right) ,\quad
\delta \sigma_2 \to {\textstyle \frac12} \left( \delta \sigma_2^0 - \delta \sigma_1^0\right) ,
\end{equation}
\begin{equation}
\delta \bar{W} \to  \delta \bar{W}^0 - \sqrt{ {\textstyle \frac{\varepsilon}{2}}} \Big( \delta \nu_1^{0}  + \delta \nu_2^{0}  + {\textstyle \frac{\nu}{3}} \delta \nu_3^{0}   \Big),
\end{equation}
while two of the spatial curvature perturbations grow
\begin{equation}
\delta \nu_1 \to \frac{ \nu}{3} \Big( \delta \nu_1^{0} - \delta \nu_2^{0} \Big) t
, \quad \delta \nu_2 \to - \frac{ \nu}{3} \Big( \delta \nu_1^{0} - \delta \nu_2^{0} \Big) t,
\end{equation}
driving the system away from the critical point, which is therefore {\it unstable}.

\section{Summary and discussion}
\label{sec_conclusion}

The combination of numerical relativity simulations in Sec.~\ref{sec_results} and the critical-point analysis in the ultra-local limit in Sec.~\ref{sec:analytics} provide complementary information about the power of slow contraction to smooth and flatten spacetime.  

The numerical studies show the robustness of slow contraction in transforming spacetimes with wildly non-perturbative inhomogeneous initial conditions into homogeneous spacetimes that approach the ultra-local limit.   The analytic studies prove that the {\it only}  possible ultra-local end states are either $\Omega_m=1$ FRW or a Kasner-like universe with a combination of anisotropy and matter energy density. 

Which end point is reached depends on the initial conditions. In a series of phase diagrams exploring different combinations of shear and intrinsic curvature inhomogeneities, we have shown that for negative exponential potentials with  ${\textstyle \frac12} (V_{,\phi}/V)^2 \equiv \varepsilon \gtrsim 13$ and physically plausible initial scalar field velocity distributions (that is, at rest or downhill for all $x$), the universe is driven to an $\Omega_m=1$ FRW spacetime with $\varepsilon_{\rm eff} =\varepsilon$, as required in bouncing and cyclic models of the universe.  Similar results were found for more complicated potentials for which the condition on $V_{,\phi}/V$ is only maintained for a short interval of time.  

The phase diagrams also show that, in some cases, an initial inhomogeneity can favor eventual convergence to the $\Omega_m=1$ FRW spacetime.  For example, there are regions of Phase Diagram III that would converge to Kasner if there is no initial inhomogeneity ($f_1=a_1=a_2=0$), 
 but that are instead driven, after a few bounces, to FRW when $a_2$ is set to an even relatively small value, such as $a_2=0.01$ in Fig.~\ref{fig:7}(a).  This suggests that the basin of attraction for the KL critical point is small. 
 In other cases, the result is a mixed state that is almost entirely FRW (as measured by proper volume) interspersed with exponentially tiny Kasner-like regions  over which the ratio of matter energy density and shear vary.   The ultimate fate of these comparatively infinitesimal Kasner-like regions is an interesting academic question that is not yet resolved, but one that is not of practical relevance to cosmology because of their insignificant volume weight.  

The bottom-line of the complementary studies is that slow contraction is an even more robust smoothing and flattening mechanism than imagined based on earlier heuristic arguments or than has been shown for any other proposed cosmological smoothing and flattening process.

\section*{Acknowledgments}
We thank David Garfinkle for helpful comments and discussions. 
The work of A.I. is supported by the Lise Meitner Excellence Program of the Max Planck Society and by the Simons Foundation grant number 663083.
W.G.C. is partially supported by the Simons Foundation grant number 654561. 
F.P. acknowledges support from NSF grant PHY-1912171, the Simons Foundation, and the Canadian Institute For Advanced Research (CIFAR).  P.J.S. is supported in part by the DOE grant number DEFG02-91ER40671 and by the Simons Foundation grant number 654561.

\newpage

\appendix

\section{Numerical methods and convergence tests}
\label{sec:app-convergence}

In this Appendix, we describe our tests for numerical convergence. The key result is  that cases of interest to bouncing cosmology -- those regions of the phase diagram that completely smooth and converge to the $\Omega_m=1$ FRW dynamical attractor solution -- strongly satisfy all tests.  The same is found to hold  in cases ending with mixed states for those regions of spacetime that smooth without encountering spikes; these regions occupy almost the entire spacetime, as measured by proper volume.   The effects on convergence are also shown for the exponentially small non-smooth regions that undergo spikey behavior.

To numerically solve the system of equations detailed in Eqs.~(\ref{Neqn}-\ref{eq-S-a-Hn}), we use 2nd order accurate spatial derivatives, and a 3 step method for time integration given by the Iterated Crank-Nicolson method. The evolution equations consist of a coupled elliptic-hyperbolic system of equations, so at each sub-step of the time integration, we first solve the elliptic equation for the Hubble-normalized lapse $\mathcal{N}$ through a relaxation method, and then update the hyperbolic equations to the next Iterated Crank-Nicolson sub-step. In the simulations demonstrated above we use a grid of 1024 points, with $\Delta x = 2\pi/1024$,  with a Courant factor of 0.5. 
	
To demonstrate the convergence of our code,  the error and convergence was analyzed for a broad range of examples drawn from Phase Diagram V (Fig.~\ref{fig:10}). Here we present  the results for two representative cases:
		\begin{eqnarray}
	c = 5,~ a_1 = 0.5,~ a_2 = 0.5,~ f_0/c = 0.5, ~ Q_0 =2.4 \label{eq:parameters1}
	\end{eqnarray}
	corresponding to a simulation that smooths everywhere to FRW without encountering spikes, and  
	\begin{eqnarray}
	c = 5,~ a_1 = 0.5,~ a_2 = 0.5,~ f_0/c = 0.5, ~ Q_0 =1.9 \label{eq:parameters2}
	\end{eqnarray}
 corresponding to a simulation that ends in a mixed state with spikes. 
 
\begin{figure*}[tb!]
\begin{center}
\includegraphics[width=5.5in,angle=-0]{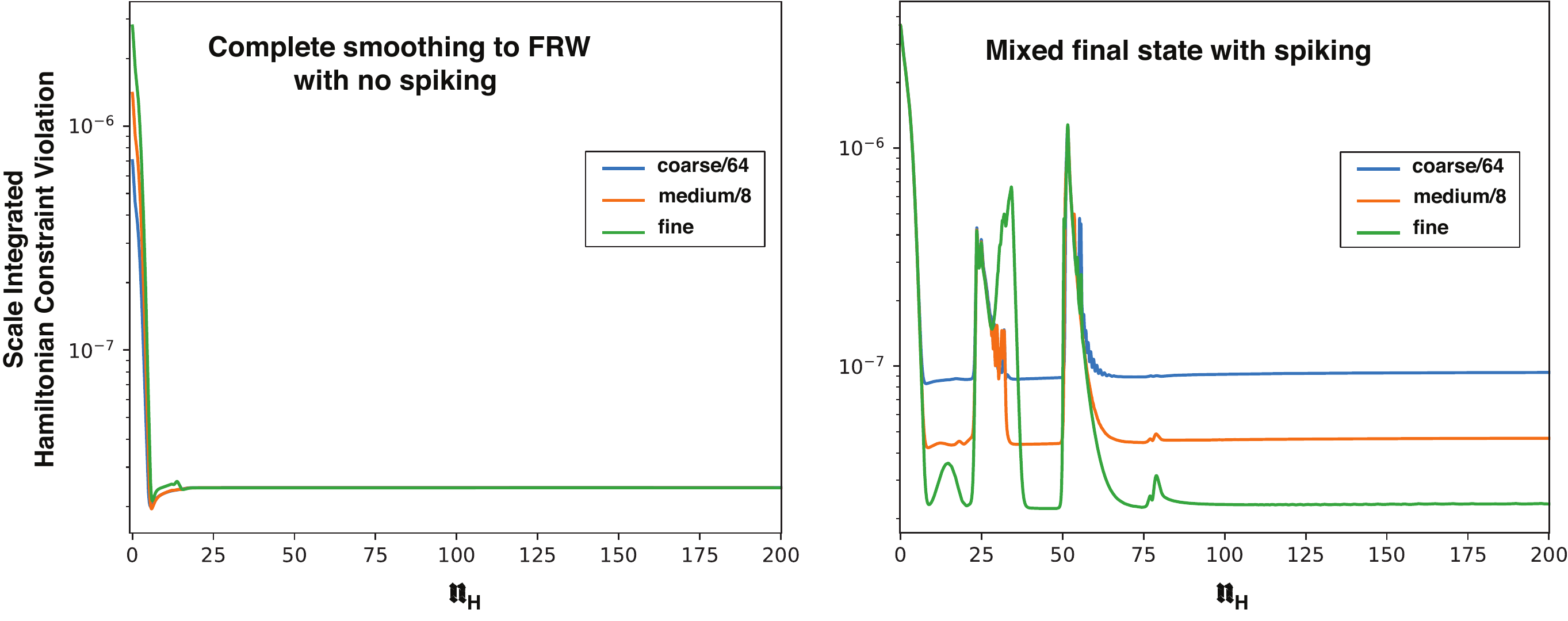}
\end{center}
\caption{	The Hamiltonian constraint integrated over the spatial domain as a function of time, for 3 resolutions, coarse  $\Delta x_c = 2\pi/512$, medium $\Delta x_m = 2\pi/1024$ and fine $\Delta x_f = 2\pi/2048$. {\it Left}, parameters of 
Eq.~\eqref{eq:parameters1}: the evolution encounters no spikes while smoothing to FRW.  We plot the integrated Hamiltonian constraint rescaled by the appropriate factor for 3rd order convergence $(||H||_c/64, ||H||_m/8, ||H||_f)$. {\it Right}, parameters of Eq.~\eqref{eq:parameters2}: the evolution does form spikes at this point. We plot the integrated Hamiltonian constraint  rescaled by the appropriate factor for 2nd order convergence $(||H||_c/16, ||H||_m/4, ||H||_f)$.    
\label{fig:A1}}
\end{figure*}   

\begin{figure*}[t]
\begin{center}
\includegraphics[width=2.75in,angle=-0]{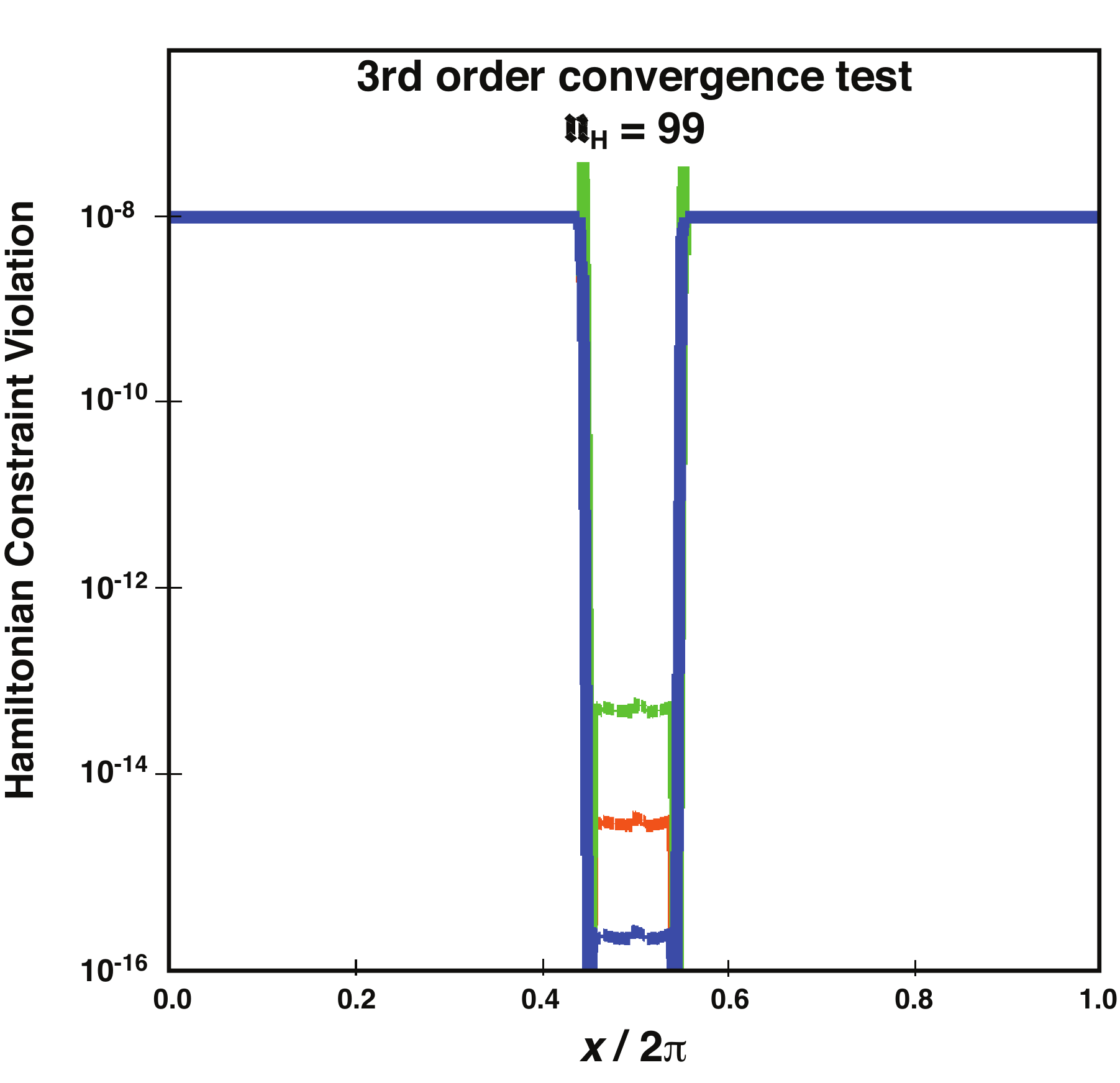}
\end{center}
\caption{The Hamiltonian constraint as a function of space at fixed times for the same 3 resolutions (with the same color coding) described in Fig.~\ref{fig:A1} (right), rescaled by factors to demonstrate 3rd order convergence, for parameters of Eq.~\eqref{eq:parameters2} that lead to a mixed final state containing an `inner region' that does not smooth to FRW surrounded by a smooth region that does. Recall that the inner region is exponentially small compared to the smoothed region  when measured by proper volume.  In the regions that smooth to FRW, the third order convergence is unaffected by the presence of the spikes in the inner region.  By contrast, convergence is lost in the inner region.    
\label{fig:A2}}
\end{figure*} 

\begin{figure*}
\begin{center}
\includegraphics[width=6.00in,angle=-0]{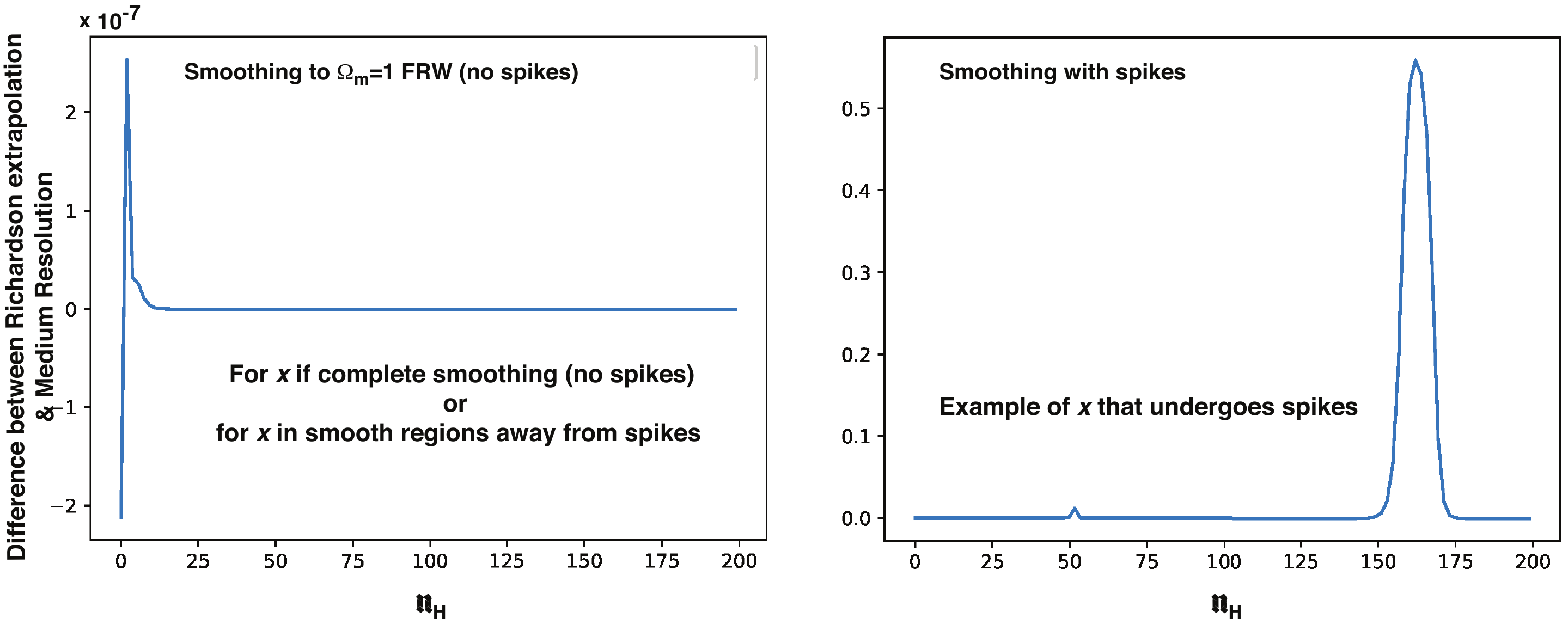}
\end{center}
\caption{The difference between the Richardson extrapolation of the trajectory of $\Sigma_+$ and the calculated value at resolution $\Delta x = 2\pi/1024$ for spatial points $x=3\pi/2$ (left) and $x=\pi$ (right), with parameters of Eq.~\eqref{eq:parameters2}. For a spatial point that smooths directly to FRW the error is consistently small, $\sim 10^{-7}$ (left). For a point that does not smooth and which remains in a Kasner-like region with spikes, the error remains small until spikes form, at which point the error grows in magnitude (right).  The growth in error corresponds to a transition between Kasner states at $x=\pi$.
\label{fig:A3}}
\end{figure*} 
	 
	  Fig.~\ref{fig:A1} shows  the L2 norm of the Hamiltonian constraint (Eq.~\ref{constraintG}) integrated over the spatial domain as a function of time. We see that, in the case of a spacetime that smooths directly to FRW everywhere, after an initial period of second order convergence, the constraint converges faster than expected, at third order (Fig.~\ref{fig:A1} left). In the case where a region of the spacetime does not smooth, we see a reduction in convergence at these points, dropping to second order or worse when spikes form. Between the times of spike formation we retain 3rd order convergence (Fig.~\ref{fig:A1} right).

For the second case, Fig.~\ref{fig:A2} shows
 the modulus of the Hamiltonian constraint as a function of the spatial coordinate $x$ at a fixed time (in this case, $n_H=99$).   Where spikes form, we see that the Hamiltonian constraint in the regions where the spikes form ($0.45\lessapprox x\lessapprox 0.55$) do not converge at the required rate; but in the outer regions, where the spacetime has smoothed to FRW without encountering spikes, the convergence properties remain unaffected.

For the same mixed state example, Fig.~\ref{fig:A3} tracks the evolution of  $\Sigma_+$ in Eq.~\eqref{Sigmaplus}  at two fixed spatial points.  The left panel focuses on a point that smooths to FRW, and the right focuses on a point that remains unsmoothed and encounters spikes. In the region that smooths to FRW, we see second order convergence for this variable as expected. Assuming second order convergence and using the data with the production resolution of 1024 grid points (medium resolution subscript $m$) and data with double the resolution (fine resolution subscript $f$), we perform a Richardson extrapolation, which is defined as	
	\begin{eqnarray}
	R = \frac{4(\Sigma_{+,f} - \Sigma_{+,m}/4)}{3}.
	\end{eqnarray}
	We take the error as the difference between this extrapolation and the medium resolution results.
	 We estimate the absolute size of the error as approximately $10^{-7}$ in the trajectory of $\Sigma_+$ that smooths to FRW (Fig.~\ref{fig:A3} left). For the trajectory not smoothing to FRW we see that the variable still converges at 2nd order until the formation of spikes, at which point errors can grow as large as $10^{-1}$ (Fig.~\ref{fig:A3}  right).

\section{Evolution of the eigensystem in the ultra-local limit}
\label{sec:app-eigensystem}

Here, we show that, in the ultra-local limit ($\bar{E}_a{}^i \to 0, \bar{A}_a \to 0, \bar{S}_a \to 0$), the eigenvectors of the shear and intrinsic curvature tensors, $\bar{\Sigma}_{ab}$ and $\bar{n}_{ab}$, at some time $t_0$ remain eigenvectors at later times.  As a result,
all the dynamics is incapsulated in the evolution of the shear and spatial curvature eigenvalues $\sigma_i, \nu_i$ ($i=1,2,3$). Of course, this is a trivial statement if the shear and intrinsic curvature tensors are diagonal. Here, we generalize to the case that $\bar{\Sigma}_{ab}$ and $\bar{n}_{ab}$ have non-zero off-diagonal components.

Since both $\bar{\Sigma}_{ab}$ and $\bar{n}_{ab}$  are symmetric and real ({\it i.e.}, Hermitian) and commute in the ultra-local limit (see Eq.~\ref{constraintC-ul}), the two tensors share a common orthogonal system of eigenvectors .

Let $\{\psi_1, \psi_2, \psi_3\}$ be an orthogonal eigensystem for   $\bar{\Sigma}_{ab}$ and $\bar{n}_{ab}$ and let $\sigma_i$ and $\nu_i$ be the eigenvalue {\color{black} corresponding} to the eigenvector $\psi_i$, {\it i.e.}, 
\begin{equation}
\bar{\Sigma}_{ab} \cdot \psi_i = \sigma_i \,\psi_i, \quad 
\bar{n}_{ab} \cdot \psi_i = \nu_i \,\psi_i, 
\quad (i=1,2,3),
\end{equation}
{\color{black} where there is no summation over $i$.}
 Then, from the evolution equations~(\ref{eq-sigma-ab-ul}-\ref{eq-n-ab-ul}), it is immediately apparent that all time derivatives can be eliminated and replaced by combinations of the original tensors $\bar{\Sigma}_{ab}$ and $\bar{n}_{ab}$. 
 It follows that $\{\psi_i\}$ are also eigenvectors of the time derivatives $\dot{\bar{\Sigma}}_{ab}$ and $\dot{\bar{n}}_{ab}$:
\begin{eqnarray}
\label{ev-der1}
\dot{\bar{\Sigma}}_{ab} \cdot \psi_i &=& - \Big( \big( 3{\cal N} - 1 \big) \bar{\Sigma}_{ab} + {\cal N} \big( 
2 \bar{n}^c{}_{\langle a} \bar{n}_{b\rangle c } - \bar{n}^c{}_c \bar{n}_{\langle ab \rangle} \big)
 \Big) \cdot \psi_i 
\equiv \lambda_i \,\psi_i
,\\
\label{ev-der2}
\dot{\bar{n}}_{ab} \cdot \psi_i &=& - \Big( \big( {\cal N} - 1 \big) \bar{n}_{ab} - 2\, {\cal N} \,  \bar{n}^c{}_{(a} \bar{\Sigma}_{b )c} \Big) \cdot \psi_i 
\equiv \eta_i \,\psi_i.
\end{eqnarray}

Using the orthogonality of the eigensystem, $\big( \psi_j \cdot \psi_i \big)=0$ {\color{black} for $i \ne j$}, and the expressions above, one can compute  $\psi_j \cdot {\rm d}_t \Big( \bar{\Sigma}_{ab} \cdot \psi_i \Big)$ in two different ways that must necessarily be equivalent:
\begin{eqnarray}
 \psi_j \cdot {\rm d}_t \Big( \bar{\Sigma}_{ab} \cdot \psi_i \Big)&=&\dot{\sigma}_i \cancelto{0}{ \Big(\psi_j \cdot \psi_i\Big)} + \sigma_i  \Big(\psi_j \cdot \dot{\psi}_i\Big) \\
 &=& \lambda_i\cancelto{0}{\Big(\psi_j \cdot \psi_i\Big) }   + \psi_j \cdot \bar{\Sigma}_{ab} \cdot \dot{\psi}_i = \sigma_j \big( \psi_j \cdot \dot{\psi}_i \big),
 \nonumber
 \end{eqnarray}
meaning that for $i\neq j$ and $\sigma_i \neq \sigma_j$, $\psi_j$ and $\dot{\psi}_i$ are orthogonal to one another, 
\begin{equation}
\psi_j \cdot \dot{\psi}_i \equiv 0,
\end{equation}
such that $\dot{\psi}_i$ is identical to $\psi_i$ up to stretching;  if all eigenvectors are simply stretched over time, then the subspace spanned by an individual eigenvector is unchanged or, equivalently, every eigenvector at time $t_0$ remains an eigenvector.  Note that each eigenvector $\psi_i$ keeps its own direction and the triad $\{\psi_1, \psi_2, \psi_3\}$ does not undergo an overall rotation.
The same argument can be repeated for $\bar{n}_{ab}$ and its eigenvalues $\nu_i$.

It remains to discuss the degenerate case. We note that the shear tensor is defined to be trace-free ($\sigma_1 + \sigma_2 + \sigma_3 = 0$) and must therefore have at least two different eigenvalues, unless all $\sigma_i \;(i=1,2,3)$ are zero.  So the only non-trivial degenerate case is one in which  two eigenvalues are the same and one is different. 
Without loss of generality, let us assume that $\sigma_1=\sigma_2$ and $\sigma_3\ne \sigma_{1,2}$ and that 
$\psi_1$ and $\psi_2$ are two linearly independent eigenvectors with eigenvalue $\sigma_1$ at time $t_0$.  Together, $\psi_1$ and $\psi_2$ span the two-dimensional subspace of eigenvectors with eigenvalue $\sigma_1$.  
 In this case, the same sort of argument as above shows that the time-evolution of $\psi_1$ and $\psi_2$ maps them into the same two-dimensional subspace, 
 although in general they could be stretched and rotated.  Furthermore, the time-evolved  $\psi_1$ and $\psi_2$ and every linear combination thereof have the same eigenvalue.  This can be seen  from Eqs.~(\ref{ev-der1}-\ref{ev-der2})  by Taylor-expanding the shear and intrinsic curvature tensors around the initial time $t_0$ and operating on an arbitrary linear combination of the $t=t_0$ eigenvectors $\psi_1$ and $\psi_2$:
\begin{eqnarray}
\bar{\Sigma}_{ab} (t_0+\Delta t)\cdot (a \psi_1 + b \psi_2) &\simeq& \bar{\Sigma}_{ab} (t_0) \cdot (a \psi_1 + b \psi_2) + \dot{\bar{\Sigma}}_{ab} (t_0) \Delta t \cdot (a \psi_1 + b \psi_2)\\
&=& \big(\sigma_1   + \lambda_1 \Delta t \big) \cdot (a \psi_1 + b \psi_2),
\nonumber
\end{eqnarray}
where we have used the fact that Eqs.~(\ref{ev-der1}) imply $\lambda_1$ must equal $\lambda_2$.  As in the non-degenerate case, every eigenvector at time $t_0$ remains an eigenvector. A similar analysis applies to $\bar{n}_{ab}$.  

Accordingly, the dynamics in the ultra-local limit is completely determined by the time-evolution of the eigenvalues.


\section{Linearized evolution equations in the ultra-local limit}
\label{sec:appPerturbedEqs}

In Sec.~\ref{sec:analytics}, we identified the end states that we found numerically as the only dynamical attractors, {\it i.e.} critical points of the evolution scheme in the ultra-local limit, given by Eqs.~(\ref{eq-sigma1-ul}-\ref{eq-w-ul}). There, we only listed the equations as linearized around the three critical points that we previously identified and listed in Table~\ref{table-1}. Here, we present the perturbed system as linearized for an arbitrary background solution, which underlied our calculations. (We shall denote perturbed quantities by $\delta$, {\it e.g.}, the $i$th linearized shear eigenvalue is given by $\delta\sigma_i$. All other variables are background quantities.)

The linearly perturbed system of ordinary differential equations describing the dynamics in the ultra-local limit around an arbitrary background solution takes the following form:
\begin{eqnarray}
\delta \dot{\sigma}_1 &=&   \Big(1 - 3{\cal N} \Big) \delta \sigma _1  
-  \left( 3  \sigma _1 +  {\textstyle \frac13 } \big(2\nu_1 - \nu_2 - \nu_3 \big) \nu_1 - {\textstyle \frac13 } \big(  \nu_2- \nu_3  \big)^2 \right) \delta {\cal N}
\\
&-&  {\textstyle \frac13 }  \, {\cal N} \Big( 4\nu_1 - \nu_2  - \nu_3 \Big) \delta \nu_1 
+  {\textstyle \frac13 }  \, {\cal N} \Big( \nu_1 + 2 \big(  \nu_2- \nu_3  \big) \Big) \delta \nu_2
+ {\textstyle \frac13 }  \, {\cal N} \Big( \nu_1 - 2 \big(  \nu_2- \nu_3  \big)  \Big)  \delta\nu_3  
,\nonumber\\
\delta \dot{\sigma}_2 &=&   \Big(1 - 3{\cal N} \Big) \delta \sigma _2 
 -  \left( 3  \sigma _2 +  {\textstyle \frac13 } \big(2\nu_2 - \nu_1 - \nu_3 \big) \nu_2 - {\textstyle \frac13 } \big(  \nu_1- \nu_3  \big)^2 \right) \delta {\cal N}
\\
&-&  {\textstyle \frac13 }  \, {\cal N} \Big( 4\nu_2 - \nu_1  - \nu_3 \Big) \delta \nu_2 
+  {\textstyle \frac13 }  \, {\cal N} \Big( \nu_2 + 2 \big(  \nu_1 - \nu_3  \big) \Big) \delta \nu_1
+ {\textstyle \frac13 }  \, {\cal N} \Big( \nu_2 - 2 \big(  \nu_1 - \nu_3  \big)  \Big)  \delta\nu_3  
,\nonumber\\
\label{nu1-ev-ul-app-p}
\delta\dot{\nu}_1 &=& \Big( 1 + {\cal N} \big(   2   \sigma_1 - 1 \big) \Big) \delta \nu_1  
+     \big(   2   \sigma_1 - 1 \big) \nu_1 \delta {\cal N}  + 2\, {\cal N} \nu_1   \delta \sigma_1  
,\\
\delta\dot{\nu}_2 &=& \Big( 1 + {\cal N} \big(   2   \sigma_2 - 1 \big) \Big) \delta \nu_2  
+     \big(   2   \sigma_2 - 1 \big) \nu_2 \delta {\cal N}  + 2\, {\cal N} \nu_2   \delta \sigma_2   
,\\
\delta\dot{\nu}_3 &=& \Big( 1 - {\cal N} \big( 2\sigma_1 + 2\sigma_2  + 1 \big)   \Big) \delta \nu_3   
-  2 \Big( \sigma_1 + \sigma_2  + {\textstyle \frac12 } \Big)  \nu_3     \delta {\cal N}  
- 2\, {\cal N}  \nu_3    \Big( \delta \sigma_1 + \delta \sigma_2   \Big)  
,\quad \\
\label{eq-w-ul-p}
\delta\dot{\bar{W}}&=& -\Big( 3{\cal N} - 1 \Big) \delta \bar{W}  - \Big( 3 \bar{W} - \sqrt{2\varepsilon} \,\bar{V} \Big) \delta {\cal N}   +  \sqrt{2\varepsilon} \,{\cal N} \delta \bar{V}  
 ,
\end{eqnarray}
where $\delta {\cal N} \equiv - {\cal N}^2 \delta ({\cal N}^{-1}) $ and $\delta {\bar V}$ are given by
\begin{eqnarray}
\label{constraintG-ul-p}
\delta ({\cal N}^{-1}) &=&  {\textstyle \frac16 } \Big( \nu_1-\nu_2-\nu_3 \Big) \delta \nu_1
+ {\textstyle \frac16 } \Big( \nu_2-\nu_1-\nu_3 \Big) \delta \nu_2 
+ {\textstyle \frac16 } \Big( \nu_3 - \nu_1 - \nu_2 \Big) \delta\nu_3 
\\
&+& \Big( 2\sigma_1  + \sigma_2 \Big) \delta \sigma_1   +\Big( 2\sigma_2 + \sigma_1 \Big)\delta \sigma_2
+ \bar{W} \delta \bar{W} 
\nonumber,\\
 \delta \bar{V} &= & - 3 \delta({\cal N}^{-1}) +  2 \Big( 2\sigma_1  + \sigma_2 \Big) \delta \sigma_1   
 + 2 \Big( 2\sigma_2 + \sigma_1 \Big)\delta \sigma_2+  2\bar{W} \delta \bar{W} . 
\end{eqnarray}

 
\section{Conventional critical-point analysis using Friedmann variables}
\label{sec:appFrVar}

In Sec.~\ref{sec:analytics}, we identified the FRW scaling attractor solution as the stable end state for all physically plausible initial conditions (assuming $V_{,\phi}/V=-\sqrt{2 \varepsilon} = {\rm constant}$). This solution is widely known in cosmology as the exact solution of the Einstein-scalar field equations when assuming an exponential potential. Here, we present the conventional derivation from the cosmology literature and relate the commonly used quantities from there with our variables.

The starting point is a gravitational action involving a single scalar field $\phi$ with canonical kinetic energy and and and exponential potential that is minimally coupled to gravity:
\begin{equation}
{\cal S} =  \int d^4 x \sqrt{-g} \left ( \frac12 R - \frac12\nabla_{\alpha}\phi\nabla^{\alpha}\phi - V(\phi) \right).
\end{equation}

Evaluating the action for the flat homogeneous but anisotropic Kasner-like metric
\begin{subequations}
\begin{gather} \label{eq:gen_kasner}
ds^2=-d \tau^2+a^2(\tau)\sum_i e^{2\beta_i(\tau)} \\
{\rm where}\, \, \; \beta_1(\tau)+\beta_2(\tau)+\beta_3(\tau)=0
\label{eq:beta_constraint}
\end{gather}
\end{subequations}
and $\tau$ is the proper FRW time coordinate,
we find the following system of evolution and constraint equations
\begin{subequations}
\begin{gather}
\label{eq:mod_einstein_1}
3H^2 - {\textstyle \frac12} \left(\beta^{\prime}_1{}^2 + \beta^{\prime}_2{}^2 +\beta^{\prime}_3{}^2 \right) = \textstyle{\frac12} \phi^{\prime}{}^2 + V(\phi), 
\\
\label{eq:mod_einstein_2}
H^{\prime} + \textstyle{\frac12} \left(\beta^{\prime}_1{}^2 + \beta^{\prime}_2{}^2  + \beta^{\prime}_3{}^2 \right) 
= - \textstyle{\frac12} \phi^{\prime}{}^2,
\\
\beta^{\prime\prime}_i+3H\beta^{\prime}_i =0, \quad (i=1,2,3),
\\
\label{eq:mod_einstein_4}
\phi^{\prime\prime} + 3H \phi^{\prime}  = - V_{,\phi} .
\end{gather}
\end{subequations}
As before, prime denotes differentiation with respect to proper FRW time $\tau$. In the following, we shall eliminate $\beta_3$ using the identity~\eqref{eq:beta_constraint}.

Next, we define the dimensionless variables that are often quoted as `Friedmann variables,'
\begin{equation}
x = \frac{ \phi^{\prime}}{H}, \quad y = \frac{\sqrt{|V|}}{H} , \quad u = \frac{\beta^{\prime}_1}{H} , \quad v = \frac{\beta^{\prime}_2}{H};
\end{equation}
and the dimensionless time variable 
\begin{equation}
{\mathfrak N}_a = \ln \big(a/a_0\big), 
\end{equation}
measuring the number of $e$-folds of contraction starting from $a=a_0$. (By definition, ${\mathfrak N}$ is negative if the universe contracts and positive if it expands.)
Note that, in the ultra-local limit, the Friedmann variables $u,v$ are identical to the  shear eigenvalues $\sigma_1, \sigma_2$ and the remaining Friedmann variables $x$ and $y$ can be identified with $\bar{W}$ and $\bar{V}$, respectively.

Using the dimensionless Friedmann variables, the Hamiltonian constraint~\eqref{eq:mod_einstein_1} reduces to 
\begin{equation}
y^2 = -3 + \textstyle{\frac12} x^2 + 2 \Big( u^2 + v^2 + uv \Big),
\end{equation}
and the homogeneous system of evolution equations~(\ref{eq:mod_einstein_2}-\ref{eq:mod_einstein_4}) takes the simple form
\begin{subequations}
\begin{eqnarray}
x_{,{\mathfrak N}_a} &=& \left(x + \textstyle{ \frac{V_{,\phi}}{V}} \right) y^2 
,\\
y_{,{\mathfrak N}_a} &=&  \left( {\textstyle \frac12 \frac{V_{,\phi}}{V}} x + y^2 + 3  \right) y
,\\
u_{,{\mathfrak N}_a} &=& u \, y^2 
,\\
v_{,{\mathfrak N}_a} &=& v \, y^2 
.
\end{eqnarray}
\end{subequations}

The system admits the following critical point solutions for $V_{,\phi}/V=-\sqrt{2 \epsilon} <0$:
\begin{subequations}
\begin{eqnarray}
&&(\sqrt{6}, 0, 0, 0);   \, \,(\varepsilon=3,\, {\rm FRW})
\\
&& (- V_{,\phi}/V, \sqrt{\textstyle{\frac12}(V_{,\phi}/V)^2-3}, 0, 0); \, \,(\varepsilon>3,\, {\rm FRW})
\\
&& (\sqrt{6- 4 \big( u^2 + v^2 + uv \big) }, 0, u, v);\, 
\, ({\rm KL}).
\label{aniso-fp}
 \end{eqnarray}
\end{subequations}
Obviously, these are the same {\color{black} critical points} we found and listed in the first two rows of Table~\ref{table-1}. However, since no (homogeneous) spatial curvature is included, the third critical point as listed in the third row of Table~\ref{table-1} cannot be recovered by means of this analysis 

It is straightforward to linearize this simple autonomous system around each of the critical points and conclude that the FRW scaling solution is a dynamical attractor if $|V_{,\phi}/V|$ is sufficiently large, in agreement with our analysis in Sec~\ref{sec:analytics}. Yet, again, this stability analysis is limited by assuming a flat Kasner-like metric as given in {\color{black} Eq.~\eqref{eq:gen_kasner}.} For example, it is {\it not} possible to recover the role of the (homogeneous) spatial curvature in driving the system away from a Kasner-like and towards the FRW solution.

Notably, for an exponential potential as given in Eq.~\eqref{pot}, finding the stable critical point, $x^2= 2 \varepsilon , V(\tau) = (3-\varepsilon) H(\tau)^2, u=v=0$, immediately yields the well-known FRW scaling attractor solution given in Eq.~\eqref{FRW-scaling-sol} after a series of simple integrations, using $x^2/2=d (H^{-1})^{\prime}$.

\bibliographystyle{plain}
\bibliography{bib_long_paper}

\end{document}